\documentstyle[pre,aps,epsf,floats,eqsecnum]{revtex}
\def\dee{\partial}     	\def\lo{\lambda_0}
\def\l{\lambda}		\def\eps{\epsilon}
\def\d{\delta}		\def\b{\beta}
\def\k{\kappa}		\def\pt{{\dee\over\dee t}}
\def\ad{\hat a^\dagger}	\def\dl{\langle\kern -.5mm\langle}
\def\bpsi{\bar\psi}     \def\dr{\rangle\kern -.5mm\rangle}
\def\bphi{\bar\phi}	\def\ginf{\rightarrow\infty}
\def\tp{t^{\prime}}     \def\a{\alpha}
\def\goto{\rightarrow}  \def\G{\Gamma}
\def\ds{\displaystyle}  
\def\bd{\hat b^\dagger}	\def\inert{\emptyset}
          \def\p{\prime}
\def\D{\Delta}          
       \def\ba{\bar a}
\def\bb{\bar b}         \def\np{n_{\phi}}
\def\L{\Lambda}         \def\exch{\leftrightarrow}
\def\bra#1{\langle #1|} \def\ket#1{|#1\rangle}
\def\lrz{\ell_{\rm rz}} \def\lf{\lambda_{\rm eff}}
     \def\br#1{\langle #1\rangle}
\def\vac{\ket 0}        \def\stnd{\bra{\>}}
  
\def\bh{\hat b}         
        
\def\ah{\hat a}

\begin{document}
\twocolumn[
\title{\bf Renormalization Group Study of the $\protect\bbox{A+B\rightarrow
\emptyset}$\/ Diffusion-Limited Reaction}
\author{Benjamin P. Lee}
\address{Institute for Physical Science and Technology, University of
Maryland, College Park, MD 20742, USA}
\author{John Cardy}
\address{All Souls College and Theoretical Physics, University
of Oxford, 1 Keble Road, Oxford OX1 3NP, United Kingdom}
\date{April 22, 1995}
\widetext\leftskip=0.10753\textwidth \rightskip\leftskip
\begin{abstract}
The $A+B\goto\inert$ diffusion-limited reaction, with equal initial densities
$a(0)=b(0)=n_0$, is studied by means of a field-theoretic
renormalization group formulation of
the problem.  For dimension
 $d>2$ an effective theory is  derived, from which the
density and correlation functions can be calculated.  We find the density
decays in time as
$a,b\sim C\sqrt{\D}(Dt)^{-d/4}$ for $d<4$, with $\D=n_0-C^\prime n_0^{d/2}
+\dots$, where $C$ is a universal constant, and $C^\prime$ is
non-universal.  The calculation is extended to the case of
unequal diffusion constants $D_A\neq D_B$, resulting in a new amplitude
but the same exponent.  For $d\le 2$ a controlled calculation is not possible,
but a heuristic argument is presented that the results above give
at least the leading term in an $\eps=2-d$ expansion.
Finally, we address reaction zones formed in the steady-state by opposing
currents of $A$ and $B$ particles, and derive scaling properties.
\end{abstract}
\bgroup\draft\pacs{{\bf KEY WORDS:}  Diffusion-limited reaction;
renormalization group; asymptotic densities}\egroup
\maketitle]

\section{Introduction}

Diffusion-limited chemical reactions are known in lower dimensions to
exhibit anomalous kinetics \cite{Kuzovkov,OTB}.  That is,
the evolution of the density depends strongly on fluctuations, and cannot
be derived from mean-field rate equations.   In this paper we apply
renormalization group (RG) techniques to the two-species reaction
$A+B\goto\inert$, with
the goal of determining systematically the effects of these fluctuations.

The model for the $A+B\goto\inert$ reaction involves two types of particles,
both undergoing diffusive random walks, and reacting upon contact to form
an inert particle.
In the density rate equation approach it is assumed that the $A$ and $B$
particles densities $a$ and $b$ are uniform, and that reactions occur at a
rate proportional to the product $ab$, giving
\begin{equation}
{da\over dt}={db\over dt}=-\Gamma ab,
\end{equation}
with rate constant $\Gamma$.  In the case of equal initial densities
$a(0)=b(0)=n_0$, the solution goes as $a,b\sim (\Gamma t)^{-1}$
asymptotically,
with an amplitude which is independent of the initial density.

It was first suggested by Ovchinnikov and Zeldovich \cite{OZ}, and
later demonstrated by Toussaint and Wilczek \cite{TW}, that relaxing
the assumption of uniformity yields a slower density decay.  In particular,
Toussaint and Wilczek made the observation that if the two species
have the same diffusion constants $D_A=D_B=D$, then the density difference
$a-b$ obeys the diffusion equation. As a result they found, by using
central limit arguments
to calculate the fluctuations in $a-b$ due to equal density, random initial
conditions, the asymptotic density\footnote{There is a misprint
in the $d=3$ amplitude of Ref.\ \cite{TW}, Eq.\ (19c).}
\begin{equation}\label{tw}
a,b\sim{\sqrt{n_0}\over\pi^{1/2}(8\pi)^{d/4}}\,(Dt)^{-d/4},
\end{equation}
where $d$ is the dimension of space.  Comparing with the
result of the rate equation, we see that for $d<4$ the asymptotically
dominant process is
the diffusive decay of the fluctuations in the initial conditions.

Using a particular version of the model,
Bramson and Lebowitz confirmed rigorously
the decay exponent of Toussaint and Wilczek, finding for $d<4$
\begin{equation}
a,b\sim C_d\sqrt{n_0}\>t^{-d/4},
\end{equation}
where $C_d$ is some constant which depends on the dimension $d$
\cite{BLi,BLii}.  In their treatment they demonstrated that the two
species are asymptotically segregated for $d<4$.  This segregation
was assumed in reference \cite{TW} in deriving Eq.\ (\ref{tw}).

Numerical simulations have confirmed the value $-d/4$ for the decay exponent
in one \cite{TW,KRi}, two \cite{TW,CDCii}, and three dimensions
\cite{Leyvraz}.   For all of these simulations restrictions were placed on the
occupation number per site, and usually only single occupancy allowed.
In the one-dimensional simulation of Toussaint and
Wilczek the initial density was varied, and reasonable agreement was found
with their analytic result, Eq.\ (\ref{tw}) \cite{TW}.   However, in
higher dimensions the
$\sqrt{n_0}$ amplitude dependence, when tested, has not been observed
\cite{CDCii,Leyvraz}.  In the former case the initial average occupation
number per site was kept low, whereas for the higher dimensional simulations
it was necessary to start with a nearly full lattice in order to reach the
asymptotic regime.  This suggests that Eq.\ (\ref{tw}) might not be a
universal result, but rather a limit for small initial density $n_0$.

While $d=4$ appears to be the upper critical dimension for homogeneous
initial conditions, this is not the case when the two species are
initially segregated, where instead the upper critical dimension is found to
be $d=2$ \cite{CD}.  That is, as a
result of the segregation, a localized region forms in which nearly all
reactions occur.   This reaction zone exhibits scaling behavior, and
the characteristic
exponents are independent of the dimension $d$ when $d>2$, but crossover
to dimension-dependent values for $d<2$.  Hence, one of our goals
in applying RG techniques is to better understand the role of the
dimensions $d=2$ and $d=4$.

The problem can be mapped to a field theory by starting from a master
equation description of the model \cite{PelitiRev}.  From an analysis of
the field theory we find that there is an upper critical dimension
$d_c=2$, which is associated with the stochastic processes of reaction
and diffusion.  Hence, for $d>2$
one can replace the full field theory with an effective theory, which is
valid for asymptotically late times, while for $d\le 2$ one must instead
perform an explicit renormalization group calculation.

The effective theory for $d>2$ is equivalent
to the deterministic partial differential equations
\begin{equation}\label{pde}
\dee_t a=D_A\nabla^2 a-\Gamma ab\qquad\dee_t b=D_B\nabla^2 b-\Gamma ab
\end{equation}
with stochastic, non-negative  {\it effective} initial conditions.
In deriving the effective theory we find that the initial distribution
is finitely renormalized due to the presence of relevant initial
terms, the analog of surface terms for a $t=0$ boundary in a $d+1$ dimensional
theory.  The resulting distribution can be characterized by a parameter
$\Delta$ which depends {\it nonuniversally} on the initial density.

We demonstrate explicitly that from these  equations
follows generally the asymptotic segregation of the $A$, $B$ particles
when $d<4$, and subsequently the universal decay exponent $-d/4$.
However, the amplitude of the density decay depends on the initial conditions,
and is therefore nonuniversal.  It is important
to note that if one uses instead central limit arguments to calculate the
initial distribution which should be fed into (\ref{pde}),
 then one is implicitly making the assumption that these
equations hold for all times, rather than just asymptotically.  Such an
assumption will get the exponent correct, but we claim that it does not,
in general, predict the correct amplitude because it neglects the
dynamics at short times.

For $2<d<4$ and $D_A=D_B$ we find
\begin{equation}\label{den}
\br a,\br b\sim{\sqrt{\D}\over\pi^{1/2}(8\pi)^{d/4}}\,(Dt)^{-d/4},
\end{equation}
where the angular brackets denote averages over both the processes of
reaction and diffusion and the initial conditions.
Here $\D$ is the coupling constant of the induced initial terms,
and can be calculated as an expansion in the initial density, giving
\begin{equation}\label{Delta}
\D=n_0-{(d+2)(d+4)\over 384(8\pi)^{d/2-1}\sin(\pi(d-2)/2)}\l_{\rm eff}^{d/2}
n_0^{d/2}+\dots,
\end{equation}
where $\l_{\rm eff}$ is a nonuniversal effective rate constant,
defined in section \ref{efftheory}  and used in (\ref{eom}).
Hence, in the small $n_0$ limit the amplitude is universal, and we recover
the result of Toussaint and
Wilczek, Eq.\ (\ref{tw}).  The
higher order  terms in $n_0$ are nonuniversal, and offer a  possible
explanation
for the deviation from $\sqrt{n_0}$ behavior
found in the simulations \cite{CDCii,Leyvraz}.

Our results (\ref{den}) and (\ref{Delta}) appear to disagree
with those of  Bramson and Lebowitz \cite{BLii},  for $d=4$ as well as
in the case above.  However, we stress that since the dependence of the
amplitude on the initial density is nonuniversal there is no
explicit contradiction.   Our model is defined by a continuous time
master equation in which the reaction occurs at a rate $\l$, and
multiple occupancy per site of each particle type is allowed.
Bramson and Lebowitz also study a continuous time model with multiple
occupancy allowed, but with an instantaneous reaction \cite{BLi,BLii}.
In this case a lattice site can only contain one type of particle.
   We use a finite reaction rate since
this is convenient for mapping to the field theory, and because it allows
one to determine better the extent of universality.
   However, we cannot directly relate our results to those
of Bramson and Lebowitz, since the field theory techniques we use are no
longer valid in the limit $\l\to\infty$, to which their model corresponds.
We note that if
our results should be valid for large but finite $\lambda$, then
$\lf$, given by (\ref{Delta}), goes
to a limiting value of the order $h^{d-2}$, where
$h$ is the short distance cutoff.

For $d\le 2$ the full field theory and the subsequent renormalization must
be considered.  We find that the field theory may be exactly renormalized,
as was shown previously by Peliti for the one-species reactions $A+A\goto A$
and $A+A\goto\inert$ \cite{Peliti}.
However, the $\eps$-expansion
calculation of observables requires non-perturbative sums over
all orders of the initial density $n_0$ and the parameter $\D$,
and while these may be carried out straightforwardly in the
one-species reaction \cite{Lee}, we are unable to apply these methods to
the present case beyond the leading order in $\eps=2-d$. Thus, at least in
this approach, we are not yet able to establish
even the power law $t^{-d/4}$ for the density to all orders in $\eps$,
although we believe it to be true.

We also consider the case of unequal diffusion constants, $D_A\neq D_B$,
and show from the effective theory that in the small $n_0$ limit
\begin{equation}
\br{a(\d)}\sim\sqrt{Q(d,\d)}\>\br{a(\d=0)}
\end{equation}
where $\d=(D_A-D_B)/(D_A+D_B)$, and
\begin{equation}
Q(d,\d)={4\Bigl[(1+\d)^{2-d/2}+(1-\d)^{2-d/2}-2\Bigr]\over\d^2(d-2)(d-4)}.
\end{equation}
Therefore this falls into the same universality class, with regard to the
decay exponent, as the symmetric case.

{}From the effective field theory for $2<d<4$ it follows that the density
difference $a-b$ is at late times a gaussian random field.  This, combined
with the asymptotic segregation $a+b\sim|a-b|$ allows one to calculate
any correlation function.  We calculate exactly the equal time two-point
correlation functions $\br{a(r)a(0)}$ and $\br{a(r)b(0)}$.

The final topic we discuss is that of reaction zones, which form whenever
$A$ and $B$ particles are segregated.  One example of a reaction zone
is that which results from opposing currents of $A$ and $B$ particles.
We apply RG methods to this steady-state case, and show that the densities
and the rate of reaction have universal scaling forms.  The upper
critical dimension for this system is $d_c=2$.  These results can be
extended to apply to reaction zones in initially segregated systems
\cite{CD,GR,BR}, and also homogeneous systems for $d<4$ \cite{LC}.

\section{The Model and the Corresponding Field Theory}
\label{model}
%
%
The model is defined by a continuous time master equation for
the probability $P(\{m\},\{n\},t)$.  The set $\{m\}$ denotes the  occupation
numbers of $A$ particles on each lattice site, $\{n\}$ the occupation
numbers of $B$ particles, and $P$ is the probability of a given configuration
occurring at time $t$.  The master equation for $P$ reads
\begin{eqnarray}\label{me}
\pt&& P(\{m\},\{n\},t)=\nonumber\\
&&{\ds D_A\over \ds h^2}\sum_{i,j}\biggl\{(m_j+1)
P(\dots,m_i-1,m_j+1\dots,t)-m_iP\biggr\}\nonumber\\
+&&{\ds D_B\over \ds h^2}\sum_{<ij>}\biggl\{(n_j+1)P(\dots,n_i-1,n_j+1
\dots,t)-n_iP\biggr\}\nonumber\\
+&&\l\sum_i\biggl\{(m_i+1)(n_i+1)P(m_i+1,n_i+1,t)-m_in_iP\biggr\},\nonumber\\
\end{eqnarray}
where $D_A$, $D_B$ are the diffusion constants for $A$ and $B$ particles,
$h$ is the size of the hypercubic lattice, and $\l$ is the microscopic
reaction rate constant.
In the first two curly bracket terms, which describe the diffusion of $A$
and $B$ particles, $i$ is summed over all sites, and $j$ is summed
over nearest neighbors to $i$.

The initial conditions for $P$ are given by a Poissonian distribution,
with the average occupation number per lattice site equal to $\bar n_0$
for each species.  That is,
\begin{equation}\label{ic}
P(\{m\},\{n\},0)=e^{-2n_0}\prod_i{\bar n_0^{m_i+n_i}\over m_i!\>n_i!}.
\end{equation}

\subsection{Mapping to Field Theory}

The first step in mapping the master equation to a field theory is to
recast it in a `second quantized' form, following a procedure developed
by Doi \cite{Doi}.  Two sets of creation and annihilation operators---$\ah$,
$\ad$ for $A$ particles and $\bh$, $\bd$ for $B$ particles---are introduced
at each lattice site.  These obey the usual commutation relations:
\begin{equation}
[\ah_i,\ad_j]=[\bh_i,\bd_j]=\d_{ij},
\end{equation}
with all other commutators zero. The vacuum ket is defined by $\ah_i\vac=0$
and $\bh_i\vac=0$ for all $i$.  In terms of  these operators the
state of the system at time $t$ is defined to be
\begin{equation}
\ket{\phi(t)}=\sum_{\{m\},\{n\}}P(\{m\},\{n\},t)\prod_i\bigl(
\ad_i\bigr)^{m_i}\bigl(\bd_i\bigr)^{n_i}\vac.
\end{equation}
The master equation can be rewritten in terms of this state as
\begin{equation}\label{sqme}
-\pt\ket{\phi(t)}=\hat H\ket{\phi(t)},
\end{equation}
with the operator
\begin{eqnarray}\label{defH}
\hat H=&&\sum_{<ij>}\biggl\{{D_A\over h^2}(\ad_j-\ad_i)(\ah_j-\ah_i)+
{D_B\over h^2}(\bd_j-\bd_i)(\bh_j-\bh_i)\biggr\}\nonumber\\
&&+\l\sum_i(\ad_i\bd_i-1)\ah_i\bh_i.
\end{eqnarray}
The formal solution of Eq.\ (\ref{sqme}) is
\begin{equation}\label{sqsol}
\ket{\phi(t)}=e^{-\hat H t}\ket{\phi(0)}.
\end{equation}

The density and other averages, which are defined in the original
occupation number representation, can be calculated from $\ket{\phi(t)}$ by
introducing the projection state
\begin{equation}\label{proj}
\stnd=\bra{0}\prod_ie^{\ah_i+\bh_i},
\end{equation}
in terms of which the average is
\begin{eqnarray}\label{average}
\dl A(t)\dr&\equiv&\sum_{\{m\},\{n\}}A\Bigl(\{m\},\{n\}\Bigr)\>
P\Bigl(\{m\},\{n\},t\Bigr)\nonumber\\
&=&\stnd\>\hat A\>e^{-\hat Ht}\ket{\phi(0)}.
\end{eqnarray}
The operator analog $\hat A$ can be derived for any $A(\{m\},\{n\})$ by
Taylor expanding the latter with respect to $m_i,n_i$, and then substituting
$m_i\goto\ad_i\ah_i$, $n_i\goto\bd_i\bh_i$.
Note that
\begin{equation}\label{projev}
\stnd\ad_i=\stnd\bd_i=\stnd,
\end{equation}
for all $i$, implying that $\hat A$ can be expressed solely in terms
of annihilation operators by first writing it in normal ordered form.

The second quantized version of the model is mapped to a field
theory by the use of the coherent state representation
\cite{PelitiRev,Schulman}.  The time evolution operator in Eq.\
(\ref{sqsol}) is rewritten via the Trotter formula
\begin{equation}\label{trotter}
\exp(-\hat Ht)=\lim_{\D t\goto 0}(1-\hat H\D t)^{t/\D t}.
\end{equation}
The right-hand side, before the $\D t\goto 0$ limit is taken, can be
regarded as a factorization of the operator into time slices, and
a complete set states inserted between each factor.  The identity is given
in the coherent state basis by
\begin{equation}\label{identity}
{\bf 1}=\int{d^2 z\over\pi}\ket z\bra z
\end{equation}
where $\ket z$ is the normalized eigenstate of the annihilation operator
with complex eigenvalue $z$:
\begin{equation}
\ket z=e^{z\ad-|z|^2/2}\vac.
\end{equation}
Equation (\ref{identity}) is generalized to a product over
all lattice sites and particle species, and then the operator $(1-\hat H\D t)$
is evaluated between successive time slices, resulting as $\D t\goto 0$
in a path integral representation of (\ref{average}).

The corresponding action is
\begin{eqnarray}\label{Sab}
S=\int d^dx\biggl[&&\int_0^{t_f}dt\biggl\{a^*\Bigl(\dee_t-
{D_A\over\bar D}\nabla^2\Bigr)a\nonumber\\
\mbox{}&&+b^*\Bigl(\dee_t-{D_B\over\bar D}\nabla^2
\Bigr)b-\lo(1-a^*b^*)ab\biggr\}\nonumber\\
&&-n_0a^*(0)-n_0b^*(0)-a(t_f)-b(t_f)\biggr].
\end{eqnarray}
where the fields $a,a^*, b, b^*$ originate from the complex variables
$z$ for each particle type.  Time has been rescaled by  the average
diffusion constant \hbox{$\bar D=(D_A+D_B)/2$}, and the coupling constant is
given by $\lo=\l h^d/\bar D$.  The time derivatives above come from the overlap
between time slices, and the other the curly brace terms
result from the operator $\hat H$.  The remaining terms are not integrated
over time and represent the random initial state, with
$n_0\equiv\bar n_0/h^d$, and the final projection state (\ref{proj}).

Averages, as defined in (\ref{average}), are given in terms of this
action by
\begin{equation}\label{FTavg}
\dl A(t)\dr={\int{\cal D}(a,a^*,b,b^*)\>A[a(t),b(t)]\>e^{-S}\over
\int{\cal D}(a,a^*,b,b^*)\>e^{-S}}.
\end{equation}
where the script ${\cal D}$ denotes functional integration.
The functional $A[a,b]$ is found by directly substituting the fields
$a,b$ for the annihilation operators $\ah,\bh$ in $\hat A$.

The time $t_f$ of the projection state is arbitrary as long $t_f>t$,
where $t$ is the time argument of the observable.  This follows directly
from the condition $\stnd\hat H=0$ for probability conservation.
The final terms can be eliminated by making the field shifts
$a^*=1+\ba$ and $b^*=1+\bb$.  Then the reaction terms are
\begin{equation}
-\lo(1-a^*b^*)ab\quad\goto\quad\lo(\ba+\bb)ab+\lo\ba\bb ab.
\end{equation}

Since the conserved mode $a-b$ plays an important role in the dynamics,
it is useful to transform (\ref{Sab}) to the fields $\phi,\bphi,\psi,\bpsi$
defined by
\begin{equation}
\phi={a+b\over\sqrt 2}\qquad\bphi={\ba+\bb\over\sqrt 2}
\qquad\psi={a-b\over\sqrt 2}\qquad\bpsi={\ba-\bb\over\sqrt 2}.
\end{equation}
The $\sqrt 2$ factors are included so that the derivative terms in (\ref{Sab})
maintain a coefficient of unity.  The subsequent action is
\begin{eqnarray}\label{S}
S=\int &&d^dx\biggl[\int dt
\biggl\{\bphi(\dee_t-\nabla^2)\phi+\bpsi(\dee_t-\nabla^2)\psi
-\d\bpsi\nabla^2\phi\nonumber\\
-&&\d\bphi\nabla^2\psi-\l_1\bphi(\phi^2-\psi^2)
-\l_2(\bphi^2-\bpsi^2)(\phi^2-\psi^2)\biggr\}\nonumber\\
&&-n_{\phi}\bphi(0)\biggr],
\end{eqnarray}
where $\d=(D_A-D_B)/(D_A+D_B)$,
the couplings are $\l_1=\lo/\sqrt 2$ and \hbox{$\l_2=\lo/4$}, and the
initial density is $n_{\phi}=\sqrt 2n_0$.  This action is the starting
point for our analysis.  Note that, since we are considering only equal
initial densities, $\br a=\br b=\br\phi/\sqrt 2$.

The mapping outlined above is a general technique, which may be applied
to many different reactions, for example, the general two-species
annihilation reaction $mA+nB\rightarrow\inert$. However, in the present
work we restrict ourselves to the case $m=n=1$.

\subsection{Diagrams and Power Counting}
\label{power}

\begin{figure}
\vskip .2in\centerline{\epsfxsize=3in\epsfbox{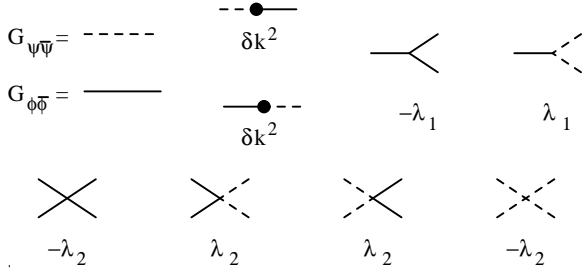}}
\vskip .15in\caption{Propagators and vertices for the full theory, given by the
action (\protect\ref{S}).}
\label{verts}
\end{figure}

A perturbation expansion for a given observable can be developed
from (\ref{FTavg}) and (\ref{S}), and expressed in
the usual diagrammatic fashion.
The propagators for $\phi,\psi$ are given by the first two terms in
(\ref{S}), and are the diffusion equation Green's function:
$G_{\phi\bphi}(k,t)=G_{\psi\bpsi}(k,t)=e^{-k^2t}$ when \hbox{$t>0$},
and $G_{\phi\bphi}=G_{\psi\bpsi}=0$ for \hbox{$t<0$}.  These
propagators are represented by solid and dashed lines respectively.  The
three- and four-point vertices, which correspond to the annihilation
reaction, are shown in Fig.~\ref{verts}.
When \hbox{$\d\ne 0$} there are also two-point vertices which connect a
$\phi$ propagator to a $\psi$ propagator, and vice versa.  These vertices
are wave number dependent, with magnitude $k^2$.

In addition there is a source term $e^{\np\bphi(t=0)}$.  By
Taylor expanding the exponential an expansion in powers of $\np$ is
generated, where the diagrams giving the $\np^i$ coefficient have a source
of $i$ $\phi$ propagators at $t=0$.  It is useful to introduce
the classical density and the classical response function, which both
involve sums over all powers of $\np$.  These quantities are important
because it is found that under renormalization $\np$ flows to a strong
coupling limit, and these sums are still meaningful in this limit
\cite{Lee}.  The term `classical' refers to absence of loops in the
diagrams.

\begin{figure}
\vskip .2in\centerline{\epsfxsize=3in\epsfbox{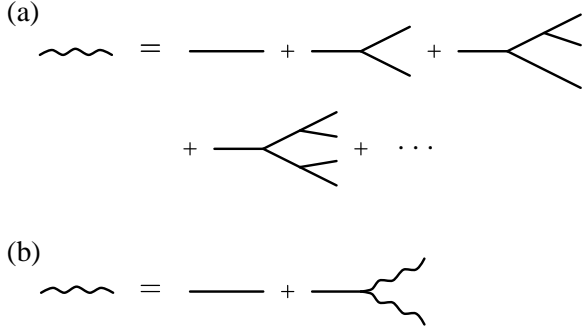}}
\vskip .15in\caption{The classical density, represented by a wavy line, is
given by (a) the sum over tree diagrams and (b) an integral equation.}
\label{treesum}
\end{figure}

The classical density is defined to be the sum of all tree diagrams which
contribute to the average $\br\phi$, as shown in Fig.~\ref{treesum}.
Note that these diagrams contain only $\phi$ propagators, because of the
three-point vertices in (\ref{S}).  This sum obeys an integral equation
which can be solved exactly, giving
\begin{equation}
\br\phi_{\rm cl}={\np\over 1+\np\l_1 t}.
\end{equation}

\begin{figure}
\vskip .2in\centerline{\epsfxsize=3in\epsfbox{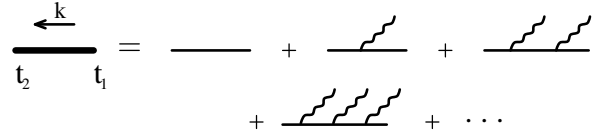}}
\vskip .15in\caption{The classical response function, shown as a heavy line,
is given
by the sum of all possible tree diagrams connected to a single propagator.}
\label{response}
\end{figure}

The classical response function is defined to be the $\phi$ propagator
with all possible tree diagrams branching off to $t=0$, as shown in
Fig.~\ref{response}.  Again this can be solved exactly, giving
\begin{equation}\label{resp}
\br{\phi({\bf k},t_2)\bphi(-{\bf k},t_1)}_{\rm cl}=e^{-k^2(t_2-t_1)}
\left(1+\np\l_1t_1\over 1+\np\l_1t_2\right)^2.
\end{equation}

In order to renormalize the field theory we must first determine the
primitive divergences, for which we consider the following dimensional
analysis.
There is a rigid constraint that $[\bphi\phi]=[\bpsi\psi]=k^d$, where
$k$ has the dimensions of wave number.  If we take the dimensions of
the conjugate fields to be $[\bphi]=[\bpsi]=k^0$, as was done for the
one-species reaction \cite{Lee}, then a general vertex
$\bpsi^i\bphi^j\psi^k\phi^\ell$ is found to be relevant only for $k+\ell\le 2$
and $d\le 2$.  Next, we observe that it is not possible to generate any
vertices with $k=1$ or $\ell=1$ from the vertices in (\ref{S}).
  Therefore all relevant
vertices are exactly those present in (\ref{S}), with an upper critical
dimension $d_c=2$.  We will discuss the renormalization of the
theory in section \ref{renormab}, and for now focus on the case for $d>2$.

To elucidate the crossover which occurs for $d>2$ it is useful to consider
rescaling the fields by dimensionful parameters, which is consistent as
long as the conjugate fields are rescaled accordingly.  Under
such a rescaling the couplings $\l_1$ and $\l_2$ behave differently, although
originally they are both proportional to $\lo$.  In particular, we can
take
\begin{equation}\label{rsc}
\phi\goto\phi/\l_1\qquad\bphi\goto\l_1\bphi\qquad
\psi\goto\psi/\l_1\qquad\bpsi\goto\l_1\bpsi,
\end{equation}
which has the result of setting the $\bphi(\phi^2-\psi^2)$ coupling to
unity, while leaving $\l_2$ unchanged.  The rescaling (\ref{rsc}) also results
in $\np\goto\l_1\np$.  This is the proper quantity to study when addressing
issues
of relevance and irrelevance, which can be seen by studying the diagrams
generated by the action (\ref{S}):  whenever an additional $t=0$ line is
added with weight $\np$, there is an additional $\l_1$ required to
connect it.  In this system of units, then, one finds that
\begin{equation}
[\l_2]=k^{2-d}\qquad[\l_1\np]=k^2\qquad[\d]=k^0.
\end{equation}
Therefore there exists a critical dimension $d_c=2$, above which $\l_2$ flows
to zero.  Doing the complete power counting method with the rescaled fields
yields the same result.
The initial density is a strongly relevant parameter for all $d$.  The
diffusion constant difference $\d$ is always marginal whenever it is not zero.

Before turning to the consequences of the power counting for the case
$d>2$, we mention
another approach to this problem, which is to integrate out the conjugate
fields $\bpsi$ and $\bphi$ in (\ref{S}).  This leads to the equations of
motion (for $\d=0$)
\begin{equation}\label{phin}
\pt\phi=\nabla^2\phi-\l_1\phi^2+\l_1\psi^2+\eta_\phi
\end{equation}
\begin{equation}\label{psin}
\pt\psi=\nabla^2\psi+\eta_\psi,
\end{equation}
where $\eta_\phi,\eta_\psi$ are multiplicative, complex noise terms
\cite{Howard}.
It is important to note that the physical density is not the field $\phi$,
but rather the average of $\phi$ over the noise terms.
These equations, without the noise terms included, are often taken as the
starting point for analysis, but this approach is not generally valid.
As we will show in the next section, one can neglect the noise terms only
for $d>2$ and for asymptotically large times.

Equation (\ref{psin}) can be simplified in any dimension, since it is a
linear equation, by averaging over the noise.  This is an
average over the stochastic process of diffusion, and not over the
initial conditions.  Then the averaged field
$\br\psi$ obeys the simple diffusion  equation for any given
initial configuration.

\subsection{Effective Field Theory for $d>2$}

\label{efftheory}
{}From the dimensional analysis and power counting above it follows that for
$d>2$ the full theory given by (\ref{S}) can be replaced by an effective
theory in which $\l_2=0$ and $\l_1\goto\l_{\rm eff}(\l_1,\l_2,\L)$,
where $\L$ is a wave number cutoff, of the order of the inverse lattice
spacing. However,
in constructing such an effective theory one has to consider all possible
relevant terms, consistent with the symmetry of the theory, which might
be generated under renormalization.
In order to identify these terms we note that this
problem is analogous to that of a semi-infinite system in equilibrium
statistical mechanics in $d+1$ dimensions,
the analog of the boundary being the
hyperplane $t=0$.  While one finds, in the semi-infinite equilibrium
case, that the bulk critical properties do not depend on the surface terms,
nonetheless one expects surface terms to contribute to correlation
functions which involve fields on the boundary \cite{Diehl}.  All
observables in our problem are
given by such correlation functions, since all diagrams originate with the
$\np\bphi(0)$ term.  Therefore we must check for all relevant {\it initial}
terms, the $t=0$ analog of the surface terms, which might be
generated, as well as those of the bulk.
As mentioned above, the only relevant bulk term is that of $\l_1$.

 The proper framework for determining which terms are relevant
is via the rescaled fields (\ref{rsc}). Therefore, for an initial term of the
type $(\D^{(m,n)}/m!n!)\bphi^m\bpsi^n\vert_{t=0}$ added to (\ref{S}) we
consider the dimensions of
the coupling $[\l_1^{m+n}\D^{(m,n)}]=k^{(n+m)(2-d)+d}$.  This power of
$\l_1$ also follows from calculating the number of vertices required to
attach a $t=0$
vertex of $\D^{(m,n)}$ to a given diagram.  These terms are relevant when
\begin{equation}\label{dnm}
d<{2(n+m)\over n+m-1}.
\end{equation}

If $m+n=1$ then the initial term is relevant for all $d$.  The case
\hbox{$m=1$} corresponds to the initial density, which has
already been demonstrated to be relevant.  For the case $n=1$ we first address
a symmetry of the theory.  When starting with equal initial densities
the system is invariant under exchanging $A\exch B$ and $D_A\exch D_B$.
Therefore the action must be invariant under the transformation
$(\phi,\bphi,\psi,\bpsi,\d)\goto(\phi,\bphi,-\psi,-\bpsi,-\d)$.
For what follows we will consider only the case $\d=0$, or
\hbox{$D_A=D_B$}, in which case the
symmetry forbids the generation of a initial term $\D^{(0,1)}\bpsi$.
In section \ref{uneqdiff} the case $\d\ne 0$ will be discussed, and it will
be demonstrated that again no $n=1$ initial term is generated.

For $m+n=2$ symmetry allows only
the generation of $\D^{(2,0)}$ and $\D^{(0,2)}$.  Below we will address
the calculation of these quantities, and demonstrate that
\hbox{$\D^{(0,2)}=-\D^{(2,0)}\equiv\D$}.  These terms are relevant
whenever \hbox{$d<4$}, as can be seen by equation (\ref{dnm}), and
therefore must be considered when constructing an effective theory
for \hbox{$2<d\le 4$}.  In fact, it will be shown
that the term $(\D/2)\bpsi^2$ is solely responsible for determining
the asymptotic decay of the density.  This is an important point.
This system is dominated by initial terms, as opposed to the one-species
reaction.  Therefore techniques which utilize
homogeneous source terms and look for a bulk steady state will not work
for this problem.  Since this initial term dominates the asymptotic
behavior of the density, we identify $d_c^*=4$ as a second critical
dimension of the system.

Higher order initial terms will also be relevant in the range $2<d\le 3$.
In fact, as $d\goto 2$ one finds that all initial terms become relevant.
While this seems to be an extreme complication, it is in fact possible
to calculate exactly the asymptotic density for $2<d\le 4$ and demonstrate
that it is independent of such terms.  This will be presented in the
next section.  We now turn to the calculation of the parameter $\D$.

\begin{figure}
\vskip .2in\centerline{\epsfxsize=3in\epsfbox{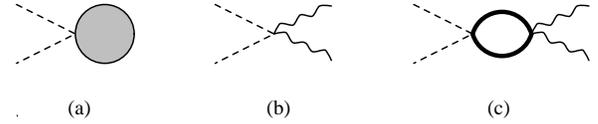}}
\vskip .15in\caption{The initial term $\D$ is generated by
diagrams of the form (a).  The tree diagrams in (b) give the leading
order contribution for small $n_0$.  The leading order corrections come
from the diagrams (c).   For $\D^{(2,0)}$ the same
diagrams would be used, but with the opposite sign for the $\l_2$ vertex on
the left.}
\label{surface}
\end{figure}

The diagrams which must be considered in calculating an effective initial
term $(\D/2)\bpsi^2$ are all those in which two $\psi$ lines exit from the
left, as shown in Fig.~\ref{surface}(a).  The sum of these diagrams
gives rise to an effective term $f(t)\bpsi(t)^2$ in the action.  If the
function $f(t)$ goes to zero for large $t$, and is sharply peaked enough
that $\int_0^\infty dtf(t)$ is finite, then a coarse-graining in time
gives $f(t)\bpsi(t)^2\to(\D/2)\d(t)\bpsi(0)^2$, where both quantities are
understood to be integrated over $t$, and $\D=2\int_0^\infty d\tp f(\tp)$.
To calculate this parameter $\D$ we consider first the subset of diagrams
given by the tree diagrams, as shown in Fig.~\ref{surface}(b).  These
diagrams sum to give $f_0(t)=-\l_2/(1+\np\l_1 t)^2$, and so
\begin{equation}
\D_0=-2\l_2\int_0^\infty dt{1\over (1+\np\l_1 t)^2}=2\np{\l_2\over\l_1}.
\end{equation}
Therefore we conclude that this set of
diagrams generates an effective initial term $\D_0=2\np\l_2/\l_1$, or,
in terms of the parameters in the master equation (\ref{me}),
$\D_0=n_0$, the initial density of each species.  This will be shown to be
the leading order term for a small $n_0$ expansion of $\D$.  The width of
the function $f_0(t)$ is given by $(\np\l_1)^{-1}$, and therefore we
expect this coarse-grained picture to be valid for times $t\gg (\np\l_1)^{-1}$.

We can group all the diagrams in the full theory (\ref{S}) which are of the
form specified in Fig.~\ref{surface}(a) in the following way.  There is
a vertex $\l_2$ which is the leftmost vertex in the diagram.  The lines
coming into this vertex from the right can either come from mutually distinct
or connected diagrams.  The tree diagrams are a subset of the former
group, and we argue that by letting $\l_1$ go to some bulk effective
coupling $\lf$ all diagrams of the former group are included.
The connected diagrams can be grouped by the number of times they are
connected, and shown in Fig.~\ref{surface}(c) are a set of diagrams which
are connected exactly once.  Again we argue that by taking $\l_1\goto\lf$
the diagrams of Fig.~\ref{surface}(c) give the entire contribution of
the set which are connected exactly once.  The sum of these diagrams is
evaluated in the appendix,
and is found to contribute to $\D$ a term which is higher order in $n_0$
than that given by the tree diagrams.  It can be shown in general that
the groups with more connections will contribute correspondingly higher
order terms, and therefore this classification scheme gives rise to an
expansion for $\D$.

Of course, an almost identical mechanism will also generate a term
$\D^{(2,0)}\bar\phi(0)^2$, and it is straightforward to show on the
grounds of symmetry that $\D^{(2,0)}=-\D$. Although this initial term is
equally relevant from the renormalization group point of view,
nevertheless it does not affect the late time behavior of the density.
This is because it acts as a source for late time fluctuations only
through the response function (\ref{resp}), and this is strongly damped
for $t_2\gg t_1$. In contrast the response function of the $\psi$ field
is simply the diffusion propagator, which has no such damping.

In summary of the discussion above, for $2<d\le 4$ and for
large times one
can replace the full theory with a simplified action
\begin{eqnarray}\label{Sf}
S=\int d^dx\biggl[&&\int_0^t dt
\Bigl\{\bphi(\dee_t-\nabla^2)\phi+\bpsi(\dee_t-\nabla^2)\psi\nonumber\\
&&-\l_{\rm eff}\bphi(\phi^2-\psi^2)\Bigr\}-n_{\phi}\bphi(0)-{\D\over 2}
\bpsi(0)^2\nonumber\\
&&+\hbox{other initial terms}\biggr],
\end{eqnarray}
where $\D$ is given by (\ref{Delta}).  Since the bulk theory, the terms
within curly braces,  is linear in $\bphi$ and $\bpsi$
these fields can be integrated out to yield the equations of motion
\begin{equation}\label{eom}
\pt\phi=\nabla^2\phi-\l_{\rm eff}\phi^2+\l_{\rm eff}\psi^2
\end{equation}
\begin{equation}\label{eomii}
\pt\psi=\nabla^2\psi.
\end{equation}
These are equations for classical fields with fluctuations in the initial
conditions.  They are often taken to be the continuum limit of the master
equation (\ref{me}), but we stress that only for $d>2$ and large times
are these equations valid. In addition, since $\l_{\rm eff}<\l$, it is
never correct to say that $\langle ab\rangle\sim\langle a\rangle\langle
b\rangle$, but only that they are effectively proportional.

\section{Density Calculation for Equal Diffusion Constants}

\subsection{Effective action: $2<d<4$}

\label{classden}
Starting with the action (\ref{Sf}) one can calculate exactly the leading
time dependence of the density, as well as correlation functions.
We begin with a comment about notation.  For this section and the next, where
we deal with only the effective field theory,  averages over the initial
conditions will be denoted by angular brackets.  The classical
fields $\phi,\psi$ themselves represent bulk averages, or equivalently,
averages over reaction and diffusion, of the
same fields as written in (\ref{S}).  Also, the effective coupling is
abbreviated to be $\l=\lf$.  With this notation, then, the average of
equation (\ref{eom}) over the translationally invariant initial conditions is
\begin{equation}\label{eomc}
{d\over dt}\br\phi=-\l\br{\phi^2}+\l\br{\psi^2},
\end{equation}
since $\nabla^2\br{\phi}=0$.

\begin{figure}
\vskip .2in\centerline{\epsfxsize=3in\epsfbox{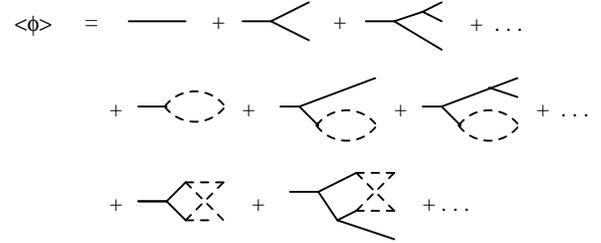}}
\vskip .15in\caption{Diagrammatic expansion for $\br\phi$.
Diagrams which contain initial terms other than $\np$
and $\D$ are not shown, but are included in the sum.  The only diagram
in which the leftmost vertex is connected to $\psi$ fields is that of
the single $\psi$ loop.}
\label{densum}
\end{figure}

A diagrammatic expansion for $\br\phi$ is shown in Fig.~\ref{densum}.
Operating on both sides of this expansion with $(\dee_t-\nabla^2)$, the
inverse of the Green's function propagator, gives equation (\ref{eomc}).
At this point, knowing that $\D$ is relevant for $d<4$, one might
attempt to apply the renormalization group to try to find a nontrivial
fixed point of order $4-d$. However, no such fixed point exists. This is
because there are no corrections to a correlation function
$\langle\psi(x_1,t_1)\psi(x_2,t_2)\bar\psi(0)^2\rangle$, so that $\D$
is not renormalized. It therefore flows, for $d<4$, to infinity under
renormalization.
This would appear to make it very difficult to sum the
diagrams in  Fig.~\ref{densum} explicitly.
However, it turns out to be possible to solve (\ref{eomc}) exactly for
late times.
There is only one diagram contributing
to the value of $\br{\psi^2}$ in equation (\ref{eomc}), which is the single
$\psi$ loop.  Evaluating this loop gives $\br{\psi^2}=\D/(8\pi t)^{d/2}$.
It is important to note that
this result holds even when all possible higher order initial terms are
included.

Next, consider a related problem in which $\br{\phi^2}$ in (\ref{eomc})
is replaced by
$\br\phi^2$, which is equivalent to including only the diagrams in
Fig.~\ref{densum} which are disconnected to the right of the leftmost vertex.
 This partial sum satisfies a differential equation known as
Ricatti's equation, which, though non-linear, can be solved.\footnote{For
an interesting presentation of the properties and history of this equation,
see \cite{Watson}.}  Let $f$ denote the  function which satisfies this
equation, that is,
\begin{equation}\label{ricatti}
{d\over dt}f=-\l f^2+\l{\D\over(8\pi)^{d/2}}
t^{-d/2}.
\end{equation}
It will be shown below that this function $f$ provides a upper bound
for the actual density, but first we will discuss the solution of this
equation.
It is integrable for certain values of $d$, specifically
$d=4$ and $d=4\pm 4/(2s+1)$ where $s$ is a non-negative integer.  For general
values of $d$ a solution can be obtained by transforming the equation
via the substitution $f=\dot u/(\l u)$, which gives
\begin{equation}
\ddot u={\l^2\D\over (8\pi)^{d/2}}t^{-d/2}u,
\end{equation}
a linear, second order equation whose solution can be expressed in terms
of confluent hypergeometric functions.  Therefore the asymptotic behavior
of $f$ is rigorously obtained, and is in fact what one naively
obtains by assuming $f\sim At^{-\a}$ and inserting it into (\ref{ricatti}):
\begin{equation}\label{UBi}
f\sim\cases{\D^{1/2}(8\pi t)^{-d/4} & $d<4$\cr  A_4^u\>t^{-1}
& $d=4$ \cr (\l t)^{-1} & $d>4$. \cr}
\end{equation}
When $d<4$ the asymptotic behavior comes from balancing the two terms on
the right hand side of (\ref{ricatti}), whereas for $d>4$ it comes from
balancing the $f^2$ and the $\dot f$ terms.  For $d=4$ all three terms
contribute, and the amplitude is
\begin{equation}\label{UBii}
A_4^u={1\over 2\l}+\sqrt{{1\over(2\l)^2}+{\D\over(8\pi)^2}}.
\end{equation}
The case of $d=4$ will be discussed in more detail in section \ref{dgefour}.
Notice that the asymptotic behavior of the solution $f$ is independent
of the initial conditions.  In fact, the initial conditions must be
specified at some $t_0>0$, since the equation is singular at $t=0$.
A natural choice for this initial time is that given by the
coarse-graining time scale of the effective initial conditions, that is
$t_0=(\np\l)^{-1}$.

Now we show that $f$ provides an upper bound for the actual density $\br\phi$.
Our method is to derive an equation for $\chi=f-\br\phi$ and show
that asymptotically $\chi\ge 0$.  Since $\phi$ is a real field in the
effective theory, then $h(t)\equiv\br{\phi^2}-\br\phi^2\ge 0$.
Equation (\ref{eomc}) can be rewritten
\begin{equation}
{d\over dt}\br\phi=-\l h(t)-\l\br\phi^2+\l\br{\psi^2},
\end{equation}
and then substituting $\br\phi=f-\chi$ gives
\begin{equation}\label{UBc}
{d\over dt}\chi=\l h+\l(\chi-2f)\chi.
\end{equation}
Assume that $\chi(t_0)=0$, that is we choose the initial condition for
$f$ such that $f(t_0)=\br{\phi(t_0)}$.  As mentioned above, the
asymptotic value of $f$ is independent of the choice of initial conditions.
Since $f$ is known to be positive for all $t>t_0$, then from
equation (\ref{UBc}) we know that $\dot\chi>0$ whenever $\chi<0$.
Now suppose that there exists some $t_1>t_0$ for
which $\chi(t_1)<0$.
Since $\chi(t)$ is a continuous function, it follows that there must
be some intermediate time $t_0<t<t_1$ for which $\chi(t)<0$ and
$\dot\chi(t)<0$.  This is in contradiction with equation (\ref{UBc}), and
therefore our assumption that there exists $\chi(t_1)<0$ for $t_1>t_0$
is false.

We can also find a lower bound for $\br\phi$ by noting that $\phi({\bf x},t)
\ge|\psi({\bf x},t)|$ at all points $({\bf x},t)$.  This is equivalent
to the statement that $a({\bf x},t), b({\bf x},t)$ are at all
points non-negative, when starting from any initial condition in which $a, b$
are everywhere non-negative.  While this result is somewhat intuitive, it can
be made more rigorous by considering the field equations (\ref{eom}),
(\ref{eomii})
expressed in terms of $a=(\phi+\psi)/\sqrt 2$ and $b=(\phi-\psi)/\sqrt 2$:
\begin{equation}\label{gx}
\pt a=\nabla^2a-\sqrt 2\l ab\qquad\qquad
\pt b=\nabla^2b-\sqrt 2\l ab.
\end{equation}
Given that the fields $a,b$ are initially everywhere non-negative, then for
the fields to have a negative value at a later time $t_1$ there
must be an intermediate time $0<t_0<t_1$ for which both $a(t_0)=0$ and
$\dee_t a(t_0)<0$.  However,
in the case where $a=0$ at a single point in space, then $a>0$ locally around
the point, implying that it is a local minimum and $\dee_t a>0$.  For
a region of $a=0$ equation (\ref{gx}) gives $\dee_t a=0$ in the region and
$\dee_t a>0$ on the boundary.  Therefore the fields cannot pass through
zero, and will remain non-negative.

Since $\phi\ge|\psi|$ it follows that $\br\phi\ge\br{|\psi|}$.  At
late times $\psi$ has a normal distribution, independent of the initial
distribution, which follows from the fact that $\psi$ obeys the simple
diffusion equation (\ref{eomii}).  Therefore the asymptotic value of
$\br{|\psi|}$
can be computed directly. The asymptotic distribution of $\psi$ is given by
\begin{equation}
P[\psi(t)]\propto\exp\left\{-{\psi(t)^2\over 2\br{\psi(t)^2}}\right\},
\end{equation}
from which it follows that
\begin{equation}\label{psi}
\br{|\psi(t)|}=\sqrt{{2\over\pi}\br{\psi(t)^2}}={(2\D)^{1/2}\over
\pi^{1/2}(8\pi)^{d/4}}t^{-d/4}.
\end{equation}

Given the upper bound $\langle\phi\rangle\leq
f\sim O(t^{-d/4})$ it can be shown that
$\br\phi\sim\br{|\psi|}$, that is, that the lower bound gives exactly
the density.  Since $\br{g^2}\ge\br g^2$ for any real $g$, then
\begin{equation}
\br{\phi-|\psi|}^2\le\br{(\phi-|\psi|)^2} =\br{\phi^2}+\br{\psi^2}
-2\br{\phi|\psi|}
\end{equation}
Using again $\phi\ge|\psi|$:
\begin{equation}\label{UBd}
\br{\phi-|\psi|}^2\le\br{\phi^2}-\br{\psi^2}= -{1\over\l}\br{\dot\phi}
=O(t^{-1-d/4}).
\end{equation}
Therefore $\br\phi=\br{|\psi|}+O(t^{-1/2-d/8})$, which gives
$\br\phi\sim\br{|\psi|}$ for $d<4$.  This is actually a statement about
segregation in the system, implying that to leading order the
density of $a+b$ is the same as $|a-b|$, or equivalently, that the minority
species in each region decays faster than the majority.
For $2<d<4$ then, we find
\begin{equation}\label{ans}
\br a\sim{\D^{1/2}\over\pi^{1/2}(8\pi)^{d/4}}t^{-d/4},
\end{equation}
as stated in section I, with $\D$ given by (\ref{Delta}).
Substituting the leading order term in the expansion $\D=n_0+O(n_0^{d/2})$
then gives the result of Toussaint and Wilczek \cite{TW}.  In fact, our
method is very similar to theirs, with two exceptions.  First, they use a
central limit argument to calculate $\D$, whereas we can compute it
directly from the full field theory.  It is reassuring that the answers
agree, to leading order in $n_0$.  The other difference is that they
calculate $\br{|\psi|}$, and then hypothesize the asymptotic segregation,
saying $\br\phi\sim\br{|\psi|}$.  Starting from the effective theory
(\ref{Sf}) we have shown rigorously that these quantities are asymptotically
the same.

\subsection{Effective Action: $d\ge 4$}

\label{dgefour}
When $d=4$ the upper and lower bounds for the density from section
\ref{classden}
still hold: $\br{|\psi|}\le\br\phi\le f$.  However, it is no
longer necessarily true that \hbox{$\br\phi\sim\br{|\psi|}$}, since the
bound on the
corrections, which is of order $O(t^{-1/2-d/8})$, is the same order as the
density.  The upper bound $\br\phi\leq f\sim A_4^u/t$
is given by (\ref{UBi}) and (\ref{UBii}).
Notice that for small $\l$ or small $\D$ that $A_4^u\goto 1/\l$.  Also, when
$\l$ is large or $\D$ is large then $A_4^u\goto\D^{1/2}/(8\pi)$.
However, in the intermediate region there is a smooth crossover in
the upper bound from the $\l$ dependent asymptote to the $\D$ dependent
asymptote.

The lower bound is given by $\br\phi\ge\br{|\psi|}=A_4^\ell/t$ with
$A_4^\ell=\D^{1/2}/8\pi^{3/2}$.  For large $\D$, then, the upper and lower
bounds differ by a factor of $\sqrt\pi$.  The lower bound continues to
decrease with $\D$, and therefore is not very
useful  in the small $\D$ limit.
However, since the parameter $\D$ is dimensionless
in $d=4$ one can do a perturbative expansion for small $\D$, which
results in a better lower bound.
It follows from equation (\ref{eomc}) that the zeroth order term in this
expansion is a constant, and is in fact equal to the small $\D$ limit of
the upper bound, $\l^{-1}$.  To the next order one has
\begin{equation}
\br a={1\over\l t}+{\l\D\over t}+O(\D^2),
\end{equation}
and it is plausible to conjecture that the amplitude is monotonically
increasing with $\D$.

The amplitude given by Bramson and Lebowitz
\cite{BLii}, has the form
\begin{equation}
A_4\propto\cases{\hbox{constant}&$\D<\D_c$\cr\D^{1/2}&$\D>\D_c$,\cr}
\end{equation}
that is, the amplitude is independent of $\D$ for small $\D$.
Their result seems to be at odds with our small $\D$ calculation.  However,
as discussed in section \ref{model}, there are differences between our model
and the one they study.  Since the corrections to the small $n_0$ or $\D$
 limit are non-universal, this is a possible explanation
of  the discrepancy.

When $d>4$ then it follows from the power counting of section \ref{efftheory}
that the $(\D/2)\bpsi^2$ initial term is irrelevant.  In this
case the density is given asymptotically by $\br a\sim(\l t)^{-1}$.
The power law of the density decay is independent of the dimension
of space.  The
amplitude $\l^{-1}$ will depend on the dimension and the microscopic
details, but it is independent of initial terms, or equivalently initial
conditions.

\subsection{Renormalization for $\lowercase{d}\le 2$}

\label{renormab}
When $d\le 2$ one has to consider the full theory as given by the action
(\ref{S}), and the subsequent renormalization.
Although (\ref{eomc}) is still valid formally, since the noise in
(\ref{phin}) averages to zero, we can no longer apply the upper and
lower bounds of the previous section since, in the presence of the
imaginary noise term, $\phi$ is no longer real.

\begin{figure}
\vskip .2in\centerline{\epsfxsize=3in\epsfbox{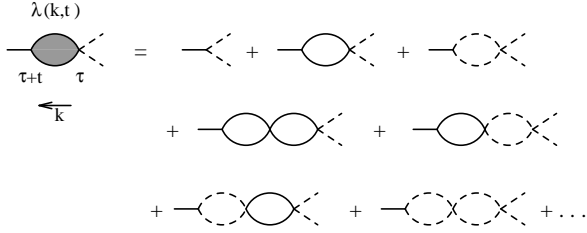}}
\vskip .15in\caption{The sum of diagrams which contribute to the
primitively divergent vertex function $\l_i(k,t)$.  Shown here is the
case $i=1$, with the $\psi$ propagators for the incoming external legs.}
\label{bubbles}
\end{figure}

Much of the contents of
this section is directly related to the one-species calculation of
Ref.~\cite{Lee}, in which more details can be found.
The primitively divergent vertex functions were identified by power counting
in section \ref{power}, and were found to be those with two lines coming in
and two or fewer lines going out.  These primitive divergences are used
to define a renormalized coupling, following conventional RG methods
\cite{Amit}.  It is found that all the vertices in the action (\ref{S})
renormalize identically, with the primitive divergences given by the
bubble sums shown in Fig.~\ref{bubbles}.

In this sum all diagrams of a given number
of loops come in with the same sign, since replacing a $\phi$ loop
with a $\psi$ loop, for example, introduces always two negative signs
(see Fig.~\ref{verts}).  At the order of $n$  loops  there are $2^n$
diagrams, so these form a geometric sum, with the ratio given by $2$
times the value of a single loop.   Denoting this sum by
$\l_i(k,t)$ where $i=1,2$ labels the number of outgoing lines, then
the Laplace transform, $\l_i(k,s)=\int_0^\infty dt\>e^{-st}\l_i(k,t)$ is
given by
\begin{equation}
\l_i(k,s)={\l_i\over 1+[4/(8\pi)^{d/2}]\,\l_2\,\Gamma(\eps/2)(s+k^2/2)},
\end{equation}
where $\eps=2-d$.
The renormalized coupling is defined in terms of an arbitrary normalization
scale $\k$, which has dimensions of wave number: $g_R=\k^{-\eps}
\l_2(k=0,s=\k^2)$.  Then the $\beta$ function is
\begin{equation}
\beta(g_R)=\k{\dee\over\dee\k}g_R=-\eps g_R+{4\eps\over(8\pi)^{d/2}}
\Gamma\left(\eps\over 2\right)g_R^2
\end{equation}
which gives a fixed point $g_R^*=O(\eps)$.

Let the density $n(t)=\br{a(t)}=\br{b(t)}$.  Since the density is
independent of the normalization scale, then $dn/d\k=0$, which leads
to the Callan-Symanzik equation
\begin{eqnarray}
\biggl[2t\pt-&&d\np{\dee\over\dee\np}-d\D{\dee\over\dee\D}
+\beta(g_R){\dee\over\dee g_R}+d\biggr]n(t,g_R,\np,\D)\nonumber\\
&&=0.
\end{eqnarray}
The solution is found by the method of characteristics to be
\begin{equation}\label{CSsol}
n(t,g_R,\np,\D)=(\k^2t)^{-d/2}n(\k^{-2},\tilde g_R,\tilde\np,\tilde\D),
\end{equation}
where in the asymptotic limit of large $\k^2t$ the running coupling
has the limit
$\tilde g_R\rightarrow g_R^*$.  However, the running initial couplings go as
$\tilde\np=\np(\k^2t)^{d/2}$ and $\tilde\D=\D(\k^2t)^{d/2}$,
that is, they flow to a strong coupling limit.

The solution (\ref{CSsol}) is used to calculate the asymptotic density
in the following way:
the density is calculated as an expansion in $g_R$ and $\np$, and
this expansion is put into the right-hand side of (\ref{CSsol}).  Then,
in the limit of large $\k^2t$,
the coupling expansion will yield an $\eps$ expansion, but only if the
behavior at large $\tilde\np$ and $\tilde\D$
is controlled.  This may be done if the diagrams may be grouped
into sums over all powers of $\tilde\np$ and $\tilde\D$, which, when summed,
yield
a well-defined limit. In the one-species
case this grouping was relatively simple. The series may be put in the
form of a sum of terms $g_R^n\tilde\np f_n(g_R\tilde\np)$, where $n$
counts the number of loops in a given diagram. The term $n=0$
corresponds to the sum of tree diagrams, given by the solution of the
simple rate equation, so that $f_0(g_R\tilde\np)\propto
(g_R\tilde\np)^{-1}$ as $\tilde\np\to\infty$. By explicit calculation,
it is then possible to show that the $f_n$ for $n>0$ behave in a similar
manner. Since $g_R\to g_R^*=O(\eps)$, this lead to the result that
$n(t)\sim A/t^{d/2}$ where the amplitude $A$ is in principle calculable
to any order in $\eps$. In the present case, the series may similarly
be organized as a sum of terms of the form
$g_R^n\tilde\np f_n(g_R\tilde\np,g_R^2\tilde\D)$, where now the
$n=0$ term is given by the sum of diagrams in  Fig.~\ref{densum}.
This is given by the solution of (\ref{eomc}), which, by the analysis of
the previous section, implies that $f_0(g_R\tilde\np,g_R^2\tilde\D)\propto
(g_R^2\tilde\D)^{1/2}/(g_R\tilde\np)$ in this limit. However, unlike the
single species case, $n$ does not simply count the loops, since already
at $n=0$ there are arbitrarily many $\psi$ loops.
In addition, while it is possible to identify those diagrams appearing
at $n=1$ for example, it is difficult to see how to express their sum
in terms of a suitable generalization of (\ref{eomc}), and thereby
evaluate it. Assuming that their asymptotic behavior is independent of
$n_0$ and thus of $\tilde\np$, there are three conceivable ways in which
these higher order terms could affect the result. They either diverge
less slowly than ${\tilde\D}^{1/2}$ as $\tilde\D\to\infty$, in which
case the $n=0$ result gives the leading behavior, which would then
yield the same result as for $d>2$; or they all behave like
${\tilde\D}^{1/2}$, in which case the density behaves as $t^{-d/4}$ but
with an amplitude which has a nontrivial expansion in powers of $\eps$;
or they diverge more strongly, in which case the density no longer
behaves as $t^{-d/4}$ for $d<2$. Since this last possibility is in
conflict with numerical experiments and rigorous results (albeit for
slightly different models), it is unlikely to be correct.

When $d=2$ the running coupling goes to zero as $(\ln t)^{-1}$ for $t\ginf$,
rather than to an order $\eps$ fixed point.  Therefore the leading order
terms for an $\eps$
expansion of the amplitude become the exact asymptotic amplitude, with
correction terms which go as $(\ln t)^{-1}$.  Therefore, if our conjecture
is correct, then density should be given exactly by (\ref{ans}) in the
large $t$ limit.

\section{Density Calculation for Unequal Diffusion Constants}

\label{uneqdiff}
When the two species of particles no longer have equal diffusion
constants, then the vertices which depend on $\d$ must be included in
the full theory.  Then for $d>2$ an effective theory can be developed,
just as before, with the resulting action
\begin{eqnarray}\label{Sun}
S=\int d^dx\biggl[&&\int_0^t dt\Bigl\{\bphi(\dee_t-\nabla^2)\phi+
\bpsi(\dee_t-\nabla^2)\psi-\bpsi\d\nabla^2\phi\nonumber\\&&
-\bphi\d\nabla^2\psi+\l\bphi(\phi^2-\psi^2)\Bigr\}-n_{\phi}\bphi(0)
-{\D\over 2}\bpsi(0)^2\nonumber\\&&+\dots\biggr].
\end{eqnarray}
The effective theory describes classical fields which evolve via the
deterministic equations of motion
\begin{equation}\label{UNa}\pt\phi=\nabla^2\phi+\d\nabla^2\psi-\l\phi^2+
\l\psi^2
\end{equation}
\begin{equation}\label{UNb}\pt\psi=\nabla^2\psi+\d\nabla^2\phi,
\end{equation}
which follows from integrating out the $\bphi,\bpsi$ degrees of freedom
in the bulk component of (\ref{Sun}).  From these equations the density can be
calculated exactly by using the same methods as before.  First, equation
(\ref{UNa}) is averaged over the initial conditions to yield equation
(\ref{eomc}),
just as in the $\d=0$ case.  The solution to Ricatti's equation again provides
an upper bound $\br\phi\leq f\sim\sqrt{\br{\psi(t)^2}}$, although the value of
$\br{\psi(t)^2}$ is changed.  It will be shown that $\br{\psi^2}\propto
t^{-d/2}$, so the upper bound decays with the same exponent as before.
Since the fields are real and $\phi\ge|\psi|$, it then follows
that $\br{\phi}\sim\br{|\psi|}$ for $d<4$, as shown in (\ref{UBd}).
Furthermore, it will be shown that
asymptotically $\psi(t)$ has a normal distribution, so the density is
given exactly by $\br a=\br\phi/\sqrt 2\sim\sqrt{\br{\psi^2}/\pi}$.
Therefore the only change in the asymptotic density from the $\d=0$ case
is due to the change in the value of $\br{\psi(t)^2}$.

\subsection{Calculation of $\br{\psi(t)^2}$}

The initial terms in the effective theory are in general changed by the
presence of $\d$ in the full theory, and therefore must
be computed again.
One can show that, as before, no $\D^{(0,1)}\bpsi$ initial term is
generated.  For any diagram which ends with a single $\psi$ line, the
last vertex (first from the left) must be a $\d k^2$ vertex.  However,
this external line has $k=0$, and so all of these diagrams have no
contribution.
To leading order $\D=n_0$ is unchanged, as can be seen from the diagrams in
Fig.~\ref{surface}:
the leading order contribution to $\D$ comes from diagrams composed of
no loops, and so all lines carry wave number $k=0$ and are unaffected by
the $\d k^2$ vertex.  The correction terms to the small $n_0$
limit of $\D$ will likely be of the same order as before,
$O(n_0^{d/2}\l^{d/2})$, but with a different amplitude.
This amplitude could be calculated, although it would require a
generalization of
the response functions discussed below.  It will be shown the asymptotic
value of $\br{\psi^2}$ depends only on $\D$, and so the other surface
terms can be neglected.

\begin{figure}
\vskip .2in\centerline{\epsfxsize=3.4in\epsfbox{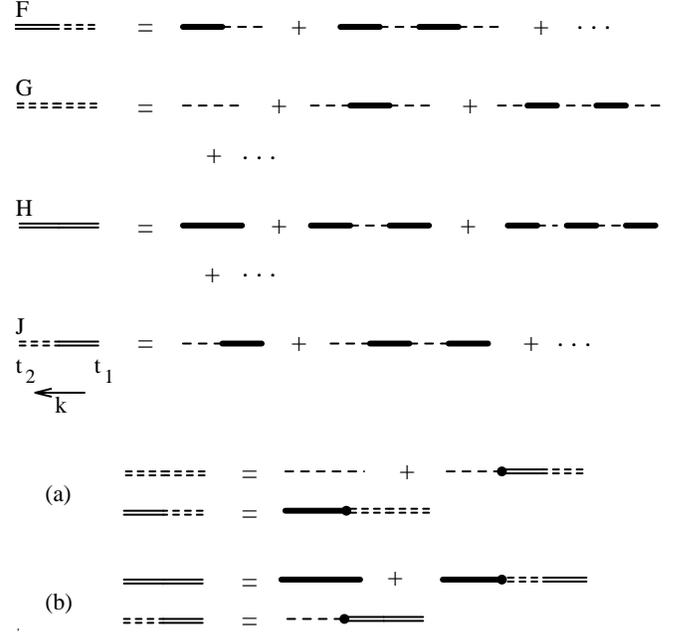}}
\vskip .15in\caption{The response functions for the case
$\d\neq 0$, and the coupled equations they satisfy.}
\label{undiffrf}
\end{figure}

There are new response
functions generated in the bulk theory.  With $\d=0$ there was just a
bare $\psi$ propagator and a $\phi$ response function.  In this
theory there are instead four response functions, which connect $\phi,\psi$ to
$\bphi,\bpsi$ in each possible way, as shown in Fig.~\ref{undiffrf}.
Each of these response functions, represented by double lines,
is an infinite sum over all possible numbers of $\d k^2$ vertices inserted.

These response functions can be found exactly via coupled integral
equations, also shown in Fig.~\ref{undiffrf}.  For our purposes, since
just the leading term for small $n_0$ is being calculated, we need only
the form of the response functions when the earlier time argument
is set to zero.  To calculate the higher order terms in the
expansion $\D=n_0+\dots$ one needs to derive these response functions
with $t_1\neq 0$.  Setting $t_2=t$, $t_1=0$ in the equations represented
by diagrams (a) gives
\begin{equation}
G(k,t)=e^{-k^2t}+\d k^2\int_0^td\tp e^{-k^2(t-\tp)}F(k,\tp)
\end{equation}
\begin{equation}
 F(k,t)=\d k^2\int_0^td\tp e^{k^2(t-\tp)}\left(1+\np\l\tp\over 1+\np\l t
\right)^2G(k,\tp),
\end{equation}
or, in terms of $f,g$ defined by $G(k,t)=e^{-k^2t}g(k,t)$ and $F(k,t)=
e^{-k^2t}f(k,t)$
\begin{equation}\label{abbot}
g(k,t)=1+\d k^2\int_0^td\tp f(k,\tp)
\end{equation}
\begin{equation}\label{costello}
f(k,t)=\d k^2\int_0^td\tp\left(1+\np\l\tp\over 1+\np\l t\right)^2g(k,\tp).
\end{equation}
Differentiating equations (\ref{abbot}) and (\ref{costello}) with respect
to $t$ gives
\begin{equation}\label{fofg}
f(k,t)={1\over\d k^2}\dot g(k,t)
\end{equation}
\begin{equation}
\pt\left[(1+\np\l t)^2f(k,t)\right]=\d k^2(1+\np\l t)^2g(k,t)
\end{equation}
Substituting for $f$ into the lower equation and manipulating
the expression gives a remarkably simple equation for $g$
\begin{equation}
{\dee^2\over\dee t^2}[(1+\np\l t)g]=\d^2k^4[(1+\np\l t)g]
\end{equation}
which has the general solution
\begin{equation}
g(k,t)={1\over 1+\np\l t}\left[A\cosh(\d k^2t)+B\sinh(\d k^2 t)
\right].
\end{equation}
{}From the integral equation (\ref{abbot}) one finds the conditions $g(k,0)=1$,
which implies $A=1$, and $g(0,t)=1$, which then implies $B=n_0\l/(\d k^2)$.
Therefore the explicit form of $G(k,t)$, and from (\ref{fofg}) $F(k,t)$, is
calculated:
\begin{equation}\label{Gpt}
G(k,t)={e^{-k^2t}\over 1+\np\l t}\left
[\cosh(\d k^2t)+{\np\l\over\d k^2}\sinh(\d k^2t)\right],
\end{equation}
\begin{eqnarray}
F(k,t)={e^{-k^2t}\over(1+\np\l t)^2}&&\biggl[\biggl(1+\np\l t-
{\np^2\l^2\over\d^2k^4}\biggr)\sinh(\d k^2t)\nonumber\\&&
+{\np^2\l^2t\over\d k^2}
\cosh(\d k^2t)\biggr].
\end{eqnarray}

The other response functions, $H(k,t)$ and $J(k,t)$, defined in diagram
Fig.~\ref{undiffrf}, can be found via similar methods.  The coupled integral
equations shown in Fig.~\ref{undiffrf}(b), written in terms of $h=e^{k^2t}H$
and $j=e^{k^2t}J$, are
\begin{equation}
h(k,t)={1\over(1+\np\l t)^2}+\d k^2\int_0^td\tp\left(1+
\np\l\tp\over 1+\np\l t\right)^2j(k,\tp)
\end{equation}
\begin{equation}
j(k,t)=\d k^2\int_0^t h(k,\tp).
\end{equation}
Differentiating both equations with respect to $t$ and substituting to
eliminate $h$ gives the equation
\begin{equation}
{\dee^2\over\dee t^2}[(1+\np\l t)j]=\d^2k^4[(1+\np\l t)j]
\end{equation}
which has the general solution
\begin{equation}\label{jpt}
j(k,t)={1\over 1+\np\l t}\left[A\cosh(\d k^2t)+B\sinh(\d k^2 t)
\right].
\end{equation}
The condition that $j(k,0)=0$ implies $A=0$.  The general solution of $h$
can be found from (\ref{jpt}), and then the condition that $h(k,0)=1$ implies
$B=1$.  Therefore $H$ and $J$ are given by
\begin{eqnarray}
H(k,t)={e^{-k^2t}\over(1+\np\l t)^2}&&\biggl[(1+\np\l t)\cosh(\d k^2 t)
\nonumber\\&&-{\np\l\over\d k^2}\sinh(\d k^2t)\biggr]
\end{eqnarray}
\begin{equation}\label{Jpt}
J(k,t)={e^{-k^2t}\over(1+\np\l t)}\sinh(\d k^2t).
\end{equation}

\begin{figure}
\vskip .2in\centerline{\epsfxsize=2.5in\epsfbox{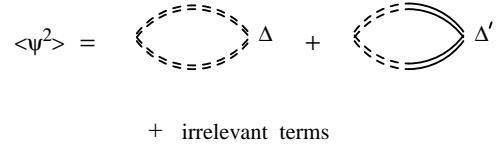}}
\vskip .15in\caption{The generalization of the simple $\psi$
loop of Fig.~\protect\ref{densum} to the case of $\d\neq 0$.}
\label{undiffloop}
\end{figure}

In section \ref{classden} the value of $\br{\psi^2}$ was calculated from the
simple loop shown in Fig.~\ref{densum}.  The generalization of this
calculation is
given by the diagrams shown in Fig.~\ref{undiffloop},
which are composed of the $G(k,t)$ and $J(k,t)$ response functions.
The surface couplings $\D^{(0,2)}\neq -\D^{(2,0)}$ beyond the leading
small $n_0$ terms, and so the couplings are labeled $\D$ and
$\D^{\p}$ respectively.
It should be noted that unlike
the $\d=0$ case, these are not the only diagrams which contribute to
$\br{\psi^2}$.  Examples of other diagrams, and arguments for why they
are irrelevant, will be given below.  First, we compute those of
Fig.~\ref{undiffloop},
which give
\begin{equation}
\br{\psi(t)^2}=\int{d^dk\over(2\pi)^d}[\D\,G(k,t)^2-\D^{\p}
J(k,t)^2].
\end{equation}
Substituting (\ref{Gpt}) and (\ref{Jpt}) into the equation above, and
rewriting the integral in terms of the variable $u=k^2t$ gives
\begin{eqnarray}
\br{\psi(t)^2}&=&{t^{-d/2}\over(4\pi)^{d/2}\G(d/2)(1
+\np\l t)^2}\int_0^{\infty}du\>u^{d/2-1}e^{-2u}\nonumber\\
&\mbox{}\times&\biggl[\D\cosh^2(\d u)-\D^{\p}\sinh^2(\d u)
+{\D\np\l t\over\d u}\sinh(2\d u)\nonumber\\
&&\mbox{}+\D\left(\np\l t\over\d u\right)^2\sinh(\d u)^2\biggr].
\end{eqnarray}
Each term in the square brackets gives a convergent integral for $d>0$.
Therefore we can take the large $t$ limit before integrating, and only
calculate the leading term in $t$, which is found to be that on the far
right in the brackets.  Consequently,  the value of $\D^{\p}$ is unimportant.

Evaluating this integral gives
\begin{equation}\label{psisq}
\br{\psi^2}={\D\over(8\pi)^{d/2}}Q(d,\d)\>t^{-d/2}
\end{equation}
where
\begin{equation}\label{Qd}
Q(d,\d)={4\left[(1+\d)^{2-d/2}+(1-\d)^{2-d/2}-2\right]\over\d^2(d-2)(d-4)}.
\end{equation}
{}From (\ref{psisq}) it follows that $Q\sim\br{\psi^2}_{\d}/
\br{\psi^2}_{0}$, in the small $n_0$ limit.  This function $Q$ is
non-singular at $\d=0$, and satisfies $Q(d,0)=1$.  While $Q$ appears
to be divergent at $d=2,4$, it is actually finite everywhere except
$d\ge 4$ and $\d=\pm 1$.  It is likely that the limits of $t\ginf$ and
$\d\goto\pm 1$ do not necessarily commute, and that a
 separate treatment for the case of an immobile species, at least in this
singular case, would be required.  For $d<4$ this function has finite
values as $\d\goto\pm 1$, but the slope at $\d=\pm 1$ is infinite for
$d\ge 2$.

\begin{figure}
\vskip .2in\centerline{\epsfxsize=3in\epsfbox{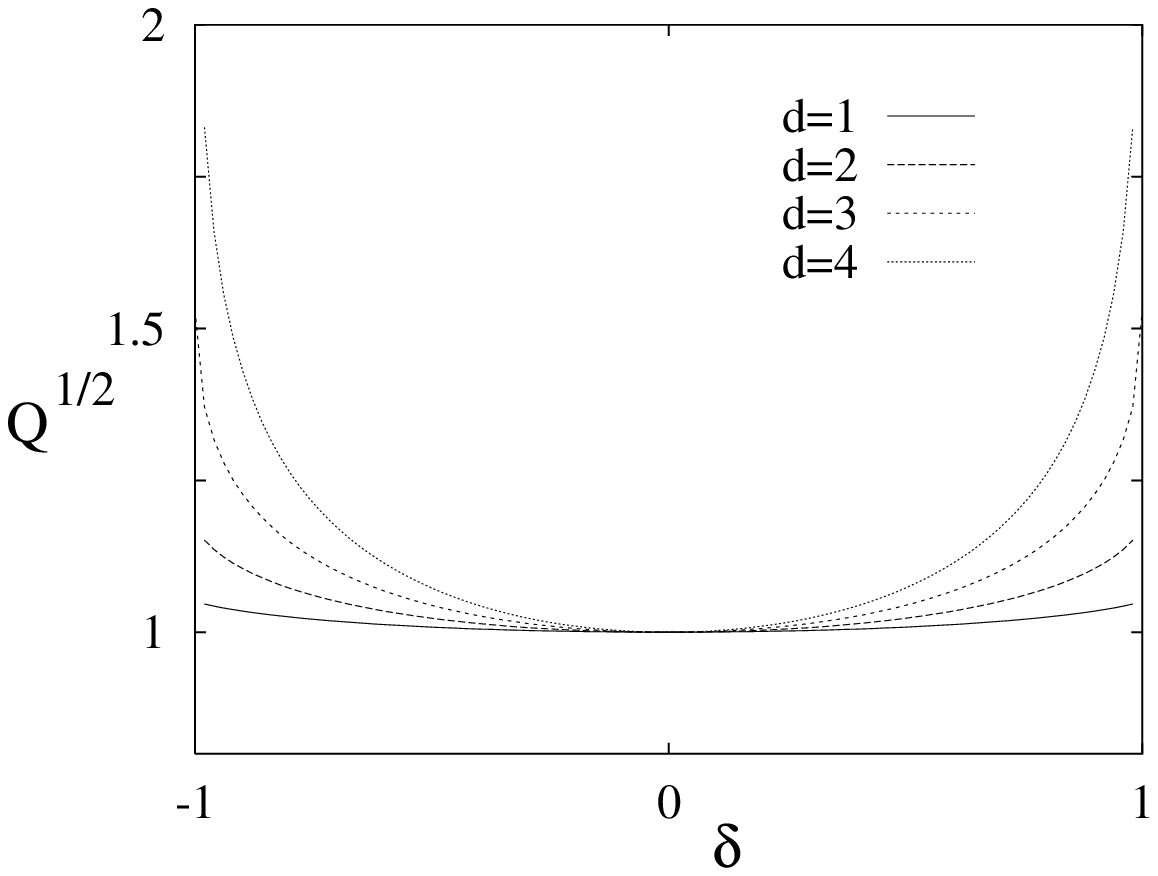}}
\vskip .15in\caption{A plot of $\protect\sqrt Q=\br a_\d/\br a_0$ for
integer values of $d$.}
\label{plotQ}
\end{figure}

While the calculation of $Q(d,\d)$ is only strictly valid for $2<d<4$, it
is nonetheless interesting to consider its limits for the integer dimensions
from $d=1$ to $d=4$, motivated by section \ref{renormab} on $d\le 2$, in
which it was conjectured that the ``classical'' amplitude is also the
leading term in an $\eps$ expansion for $d=2-\eps$.
{}From (\ref{Qd})
\begin{equation}\label{Qdii}
Q(d,\d)=\cases{
{\ds 4\over\ds 3\d^2}[(1+\d)^{3/2}+(1-\d)^{3/2}-2]&$d=1$\cr\noalign{\medskip}
{\ds(1-\d)\ln(1-\d)+(1+\d)\ln(1+\d)\over\ds\d^2}&$d=2$\cr \noalign{\medskip}
{\ds 4\over\ds\d^2}[2-\sqrt{1+\d}-\sqrt{1-\d}]&$d=3$\cr \noalign{\medskip}
{\ds -\ln(1-\d^2)\over\ds \d^2}&$d=4$\cr}
\end{equation}
Since the density goes as $\sqrt{\br{\psi^2}}$, the function $\sqrt{Q(d,\d)}$
is plotted in Fig.~\ref{plotQ}
for integer values of $d$.  The density amplitude increases monotonically
with $|\d|$, but is not changed remarkably for modest values of $\d$.

\begin{figure}
\vskip .2in\centerline{\epsfxsize=2.5in\epsfbox{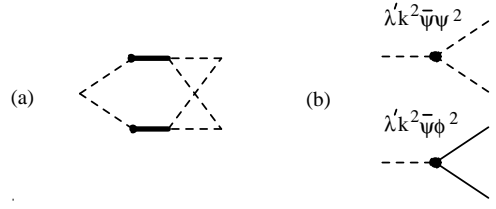}}
\vskip .15in\caption{(a) An example of one of the diagrams besides
those of Fig.~\protect\ref{undiffloop} which contribute to
$\br{\psi^2}$, and (b) the effective bulk vertices that all such diagrams
contain.}
\label{extra}
\end{figure}

There are other diagrams which give contributions to $\br{\psi^2}$,
unlike the \hbox{$\d=0$} case.  Some of these are shown in
Fig.~\ref{extra}(a).
All of these diagrams have the similar feature that they contain one of
the two sub-diagrams in Fig.~\ref{extra}(b).  These sub-diagrams give rise to
effective vertices of the form $\l^{\p}\bpsi\nabla^2\psi^2$ and
$\l^{\p}\bpsi\nabla^2\phi^2$
in the bulk theory.  However, such vertices are
irrelevant, which follows from power counting, and so the diagrams which
arise from them must be
sub-leading in time.   Therefore
we conclude that asymptotically the value of $\br{\psi^2}$ is given by
(\ref{psisq}) and (\ref{Qd}).

\subsection{Demonstration that $\psi(t)$ has a Normal Distribution}

In order for the calculation of $\br{\psi^2}$ to give the amplitude of
the density it is necessary that $\psi(t)$ have a normal distribution.
When $\d=0$ this follows directly from the simple diffusion equation
satisfied by $\psi$, or equivalently, from central limit arguments.
However, $\psi$ evolves via
equation (\ref{UNb})  for $\d\neq 0$, and so it needs to be shown that it
still flows to a normal distribution.  What we will show is that the
random variable $t^{d/4}\psi$ flows to a static normal distribution,
the width of which was calculated above.

Consider $\br{\psi^n}$, where $n$ is even.  There is one diagram in
which  $n$ response functions $G(k,t)$ are connected in pairs to $n/2$
initial terms $(\D/2)\bpsi^2$.  This diagram contains $n/2$ loops, and is
therefore of order $t^{-nd/4}$.  It was shown above replacing any of the
$G(k,t)$ loops with $J(k,t)$ response functions connected to
$(\D^{\p}/2)\bphi^2$ gives a lower order contribution.  Similarly, any other
diagrams, which would originate from considering higher order surface
terms, will involve more than $n/2$ loops, and will therefore decay
faster in time.
For $n$ odd one finds that there are no diagrams for $\br{\psi^n}$ which
decay as slowly as $t^{-nd/4}$.  That is, for $n$ odd, $\lim_{t\goto\infty}
\br{(t^{d/4}\psi)^n}=0$.  Since
the distribution of the variable $t^{d/4}\psi$ has only even moments as
$t\ginf$, and these moments are just multiples of $\br{(t^{d/4}\psi)^2}$,
generated by all possible pair contractions, then the distribution
is normal.

\section{Correlation Functions for $2<\lowercase{d}<4$}

When $d>2$, one can use the classical action to calculate the correlation
functions.
Consider the distribution of the random variable $t^{d/4}\phi({\bf x},t)$
with $2<d<4$.   From section \ref{classden} we know that
\hbox{$\br{t^{d/4}\phi-t^{d/4}|\psi|}\goto 0$} as $t\ginf$.  Furthermore,
from equation (\ref{UBd}) it follows that, as $t\ginf$, $\br{(t^{d/4}\phi
-t^{d/4}|\psi|)^2}\goto 0$.  This suggests that the distributions
$P[t^{d/4}\phi]\sim P^{\p}[t^{d/4}|\psi|]$ as $t\ginf$.
The latter distribution is known exactly, as $t^{d/4}\psi$ is at late
times given by a static normal distribution.

It is not correct to say that asymptotically $\phi$ and $|\psi|$
are everywhere equal, since this would imply that there are no regions in
which the densities $a$ and $b$ are both non-zero.  However, the reaction
regions, those in which both densities are non-zero, become negligibly small
for large $t$, and
 the corrections to setting $\phi$ equal to $|\psi|$ in
calculating correlation functions will be subleading in time.  Stated
another way, the leading term in both $\br{\phi_1\phi_2}$ and
$\br{|\psi_1||\psi_2|}$ is of order $t^{-d/2}$.  To this order the
two random variables $\phi$ and $|\psi|$ have identical distributions.
This is in contrast to a quantity such as $\phi^2-\psi^2$, which is
measuring a subleading term relative to $t^{-d/2}$.

We can use the property that $t^{d/4}\phi$ is given by the absolute value
of a gaussian random field to calculate correlation functions.
This is similar
to what is done is the dynamics of phase ordering, where the order
parameter field can be mapped to an auxiliary field which is assumed
to be a gaussian random field.  This analogy will be discussed further below.

\begin{figure}
\vskip .2in\centerline{\epsfxsize=2in\epsfbox{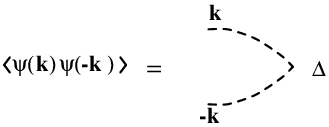}}
\vskip .15in\caption{The diagram for $\br{\psi({\bf k})\psi({\bf -k})}$,
when $\d=0$.}
\label{crossterm}
\end{figure}

Since $\phi$ and $|\psi|$ are given by the same distribution, we conclude
$\br{\phi_1\phi_2}\sim\br{|\psi_1||\psi_2|}$, where the labels indicate
the positions ${\bf x}_1$ and ${\bf x}_2$ at time $t$.
The correlation function $\br{|\psi_1||\psi_2|}$ can be calculated
exactly by using the fact that, asymptotically, $\psi(t)$ has a normal
distribution.  The joint probability distribution $P[\psi_1,\psi_2]$ is then
also normal, so
\begin{equation}\label{probdist}
P[\psi_1,\psi_2]={\sqrt{4\a^2-\b^2}\over2\pi}\exp\left\{
-\a\psi_1^2-\a\psi_2^2-\b\psi_1\psi_2\right\},
\end{equation}
where we have used translational invariance to set $\br{\psi_1^2}=
\br{\psi_2^2}$.  The constants $\a$ and $\b$ are determined by the
values of $\br{\psi^2}$ and $\br{\psi_1\psi_2}$, which are evaluated
from the diagrams.  The latter we have only calculated for $\d=0$, or
equal diffusion constants, so we consider that case first.
For notational convenience we define
$\br{\psi^2}\equiv C(t)=\D/(8\pi t)^{d/2}$.
The diagram shown in Fig.~\ref{crossterm}(a) is used to calculate
the correlation function $\br{\psi({\bf k})\psi({\bf -k})}$, from which
one finds
\begin{equation}
\br{\psi_1\psi_2}=\int {d^dk\over(2\pi)^d}e^{i{\bf k}\cdot({\bf x}_1-
{\bf x}_2)}\br{\psi({\bf k})\psi({\bf -k})}.
\end{equation}
When $\d=0$ then $\br{\psi({\bf k})\psi({\bf -k})}=\D e^{-2k^2t}$,
and
\begin{equation}\label{Deff}
\br{\psi_1\psi_2}=C(t)\exp(-r^2/8t)\equiv C(t)f(r^2/t)
\end{equation}
where $r=|{\bf x}_1-{\bf x}_2|$.  In terms of (\ref{Deff}) we find for $\a,\b$
\begin{equation}
\a={1\over 2C(1-f^2)}\qquad\qquad\b={f\over C(1-f^2)}.
\end{equation}

\begin{figure}
\vskip .2in\centerline{\epsfxsize=3in\epsfbox{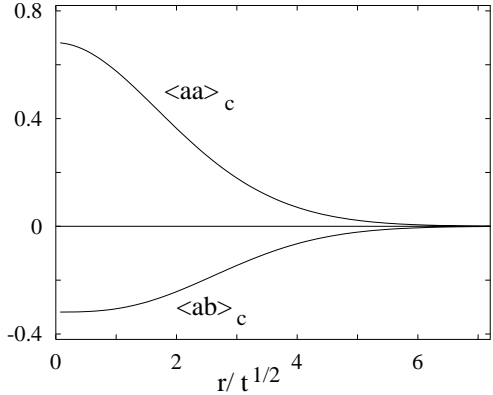}}
\vskip .15in\caption{The correlation functions $\br{a(r,t)a(0,t)}_c$
and $\br{a(r,t)b(0,t)}_c$ plotted as functions of $r/\protect\sqrt t$.
The vertical axis is given in units of $\D(8\pi t)^{-d/2}$.}
\label{corrs}
\end{figure}

With these values substituted into (\ref{probdist}), one can calculate
\begin{eqnarray}
\br{\phi_1\phi_2}&\sim&\br{|\psi_1||\psi_2|}=
\int_{-\infty}^{\infty}d\psi_1\int_{-\infty}^{\infty}d\psi_2\>|\psi_1|
|\psi_2|\,P[\psi_1,\psi_2]\nonumber\\ \noalign{\medskip}
&=&{2C\over\pi}\left[\sqrt{1-f^2}+f\arctan\left(f\over\sqrt{1-f^2}\right)
\right].
\end{eqnarray}
This correlation function can be used to find the correlation functions
$\br{a_1a_2}$ and $\br{a_1b_2}$.  Specifically
\begin{equation}
\br{a_1a_2}={1\over 2}\br{\phi_1\phi_2+\psi_1\psi_2},
\end{equation}
which gives for the connected part $\br{a_1a_2}_c=\br{a_1a_2}-\br a^2$,
\begin{eqnarray}
&&\br{a_1a_2}_c=\nonumber\\
&&{\D\over\pi(8\pi t)^{d/2}}\left[{\pi\over 2}f-1+\sqrt{1-f^2}
+f\arctan\left(f\over\sqrt{1-f^2}\right)\right]\nonumber\\
\end{eqnarray}
For large $r$, $f=\exp(-r^2/8t)$ is small, giving
\begin{equation}\label{Lra}
\br{a_1a_2}_c\sim{\D\over 2(8\pi t)^{d/2}}e^{-r^2/8t}.
\end{equation}
Similarly, $\br{a_1b_2}=\br{\phi_1\phi_2-\psi_1\psi_2}/2$, so that
\begin{eqnarray}
&&\br{a_1b_2}_c=\nonumber\\
&&{\D\over\pi(8\pi t)^{d/2}}\left[-{\pi\over 2}f-1+
\sqrt{1-f^2}+f\arctan\left(f\over\sqrt{1-f^2}\right)\right]\nonumber\\
\end{eqnarray}
which for large $r$ goes as
\begin{equation}\label{Lrb}
\br{a_1b_2}_c\sim-{\D\over 4(8\pi t)^{d/2}}e^{-r^2/8t}.
\end{equation}
A plot of these connected correlation functions is shown in Fig.~\ref{corrs}.
The signs $\br{a_1a_2}>0$ and $\br{a_1b_2}<0$ can be
understood for short distances to be a consequence of the segregation.
Given an $A$ particle at a particular point, there is an increased probability
that a nearby particle is also an $A$, and a decreased probability
that it is a $B$.

For the case $\d\neq 0$ one has $\br{\psi^2}=C(t)Q(d,\d)$, as given by
(\ref{psisq}). The generalization of $\br{\psi({\bf k})\psi({\bf -k})}$,
shown in Fig.~\ref{crossterm}, behaves for small $k$ the same as when $\d=0$.
Therefore, for large $r$ one still has $\br{\psi_1\psi_2}=Cf$.
When this is put in the expressions for $\br{a_1a_2}$ and $\br{a_1b_2}$
one finds that the large $r$ behavior is given by (\ref{Lra}) and
(\ref{Lrb}) is unaffected by $\d\neq 0$.

While these correlation functions and other quantities can be calculated,
they ultimately rely on the stronger statement that $\psi$ is a gaussian
random field, and that $\phi\sim|\psi|$.  The topology of the domains is
determined by the random field, with the boundaries between $a$ regions
and $b$ regions given by the zeroes of $\psi$.  This topology is completely
equivalent to an analogous situation in phase ordering.
It has been suggested that in the phase ordering of a scalar order parameter
an invertible, non-linear mapping from the order parameter field to
an auxiliary field results in the latter being a gaussian random field
\cite{Mazenko}.  Usually this mapping is chosen to be the solution of a single
kink, for example the hyperbolic tangent profile.
While this method is no longer believed to be quantitatively correct
\cite{Oono}, it does provide a useful picture of the structure
of the domains.  Again, the zeroes of this gaussian random field determine
the boundaries between the equilibrated phase.

The difference between these systems lies in how correlation functions are
calculated from the random field.  In the reaction-diffusion case one
is interested in the correlation functions of the field itself,  and of
the absolute value of the field.  Neither of these quantities exhibit
remarkable behavior.  In the phase ordering one argues that
at late times the mapping between the order parameter field and the
gaussian field goes to a step function, and therefore order parameter
correlations are given by the correlations of the sign of the random
field.  These sharp boundaries give rise to more interesting features,
such as non-analytic
terms in the small $r$ limit of the correlation function, or correspondingly
power law tails for large wave number in the Fourier transform.

\section{Reaction Zones}

\label{srz}
It was shown in section \ref{classden} that for $d<4$ the particles segregate
asymptotically into regions of purely $A$ or $B$ particles.  As a result of
this segregation there exist interfaces between the two species, and all
reactions occur in the interfacial regions.  These reaction zones have
interesting scaling properties.  For example, the width of the interface
goes as $t^{\a}$ with the exponent $\a<1/2$.  Also, the nearest neighbor
distance distribution of the particles in the reaction zone is found to
have a characteristic length $\lrz$ that goes as a power of $t$,
with an exponent which  differs from that
of the bulk system, where $\br a^{-1/d}\sim t^{1/4}$.
To derive these properties we begin with a related steady-state problem.

Consider a system with a source of $A$ particles located at the boundary
$x=-L$ which maintains a fixed current $J\hat{\bf x}$, and a similar source
$-J\hat{\bf x}$ of $B$ particles positioned at $x=L$.  These opposing
currents will establish a steady-state profile, in which the average
densities will be functions of the transverse coordinate $x$.  For a given
current $J$ one can choose $L$ to be large enough that the reactions are
localized
to an interfacial region of width $w\ll L$.  In this case, it is found that
the densities in the reaction zone, where $|x|\lesssim w$, have universal
scaling forms.  Also of interest is the reaction rate $R(x)=\lo
\br{a(x)b(x)}$, which exhibits scaling, and which is used to define the
width of the reaction zone.

The power counting of section \ref{power} showed that the four-point
vertices were irrelevant for $d>2$.  Therefore $R(x)\sim\l_{\rm eff}\br a\br b$
in
the asymptotic limit---which will be shown to be the small $J$
limit---and the problem reduces to the differential equations of the
effective theory:
\begin{eqnarray}\label{feq}
\pt\br a&=\nabla^2\br a-\l_{\rm eff}\br a\br b\\
\pt\br b&=\nabla^2\br b-\l_{\rm eff}\br a\br b.
\end{eqnarray}
{}From these equations it has been shown that
\begin{equation}\label{mfscale}
R\sim J^{4/3}f(xJ^{1/3}),\qquad d>2,
\end{equation}
implicitly by G\'alfi and R\'acz \cite{GR}, and later explicitly by
Ben-Naim and Redner \cite{BR}.  From (\ref{mfscale}) one identifies the
width $w\sim J^{-1/3}$, and the characteristic length of the particle
distribution within the reaction zone $\lrz\equiv\br{a(x=0)}^{-1/d}\sim
J^{-2/3d}$.  The latter quantity is derived in Ref.~\cite{LC}.

For $d\le 2$ one does not have simply differential equations, and
the full field theory must be taken into account.  We begin by observing
that the current $J_A$ is given by $a^*\dee_x a$ in the notation of
section \ref{power}, and similarly for $J_B$.  From dimensional
analysis $[J]=k^{d+1}$.

We proceed with the renormalization of the theory, as was sketched
in section \ref{renormab}.  A normalization scale $\k$ is introduced, and
used to define the renormalized coupling $g_R$.  Since physical quantities,
such as the width, cannot depend on $\k$, then
\begin{equation}
\k{d\over d\k}w=\left[\k{\dee\over\dee\k}+\beta(g_R){\dee\over\dee g_R}
\right]w(J,g_R,\d,\k)=0.
\end{equation}
Note that, since there are no diagrams which can dress the two-point
vertices in (\ref{S}), $\d$ does not get renormalized, and therefore does
not appear in equation above.  From dimensional analysis one has
\begin{equation}
\left[\k{\dee\over\dee\k}+(d+1)J{\dee\over\dee J}+1\right]w(J,g_R,\d,\k)=0.
\end{equation}
Combining these equations gives the Callan-Symanzik equation
\begin{equation}
\left[(d+1)J{\dee\over\dee J}-\beta(g_R){\dee\over\dee g_R}+1\right]w=0.
\end{equation}
with the solution
\begin{equation}\label{rzCS}
w(J,g_R,\d)=\k J^{-1/(d+1)}w(\k^{d+1},\tilde g_R,\d).
\end{equation}
In the small $J$ limit then $\tilde g_R\goto g_R^*$, and the right-hand
side is given by
\begin{equation}
w\sim J^{-1/(d+1)}f(\d,\eps),\qquad d<2.
\end{equation}
Following the same procedure for any dimensionful quantity results in the
scaling behavior being given by dimensional analysis.  That is, $\lrz\sim w$,
\begin{equation}
\br a,\br b\sim J^{d/(d+1)}F_{a,b}(x\,J^{1/(d+1)}),
\end{equation}
and
\begin{equation}
R(x)\sim J^{(d+2)/(d+1)}G(x\,J^{1/(d+1)}).
\end{equation}

Note that these results imply that \\$R\sim J^{(2-d)/(d+1)}\br a\br b$.
This can be shown explicitly by calculating $R\propto\br{ab}=\l(J)\br a\br b$,
where $\l(J)$ is given by the bubble sum shown in Fig.~\ref{bubbles},
with $s=J^{2/(d+1)},k=0$.  In the small $J$ limit then $\l(J)\sim
J^{(2-d)/(d+1)}$.

Since the current $J$ may be thought of as being due to localized
sources of $A$ and $B$ particles at $x=\pm L$ respectively, the
coupling constant
power counting arguments are formally the same as those
of Ref.~\cite{Lee} (see Sec.~\ref{renormab}),
in which the sources are
localized at $t=0$.\cite{HC}
Thus the various scaling
functions above may, in principle, be calculated as an expansion in
$g_R^*=O(\eps)$, in which the leading term is given by the solution of
the rate equations (\ref{feq}). The next order corrections to the reaction
profile have been computed in Ref.~\cite{HC}, where it was shown that the
fluctuation corrections lead to a universal power law tail in this
function.

For $d=2$ one has $\tilde g_R\sim 1/|\ln J|$ for
small $J$, and the leading order result is therefore found by
substituting this behavior into the solution of the
rate equation (\ref{feq}). This leads to the results\footnote{These logarithms
were incorrectly omitted from Ref.~\cite{LC}.}
\begin{eqnarray}
w&\propto& \left(|\ln J|\over J\right)^{1/3},\\
R(x)&\sim&\lambda(J)\br a\br b\nonumber\\
&\sim& J^{4/3}|\ln J|^{-1/3}G(x\,J^{1/3}|\ln J|^{-1/3}),\\
\lrz&\propto& J^{-1/3}|\ln J|^{-1/6}.
\end{eqnarray}

As was discussed in Ref.~\cite{LC}, the corresponding results for the
time-dependent cases of segregated initial conditions or of randomly
homogeneous initial conditions is given by substituting $J\propto
t^{-1/2}$ or $J\propto t^{-(d+2)/4}$ (with $d=2$), respectively, in the
above formulas.  These results, for the case of segregated
initial conditions, have been obtained recently via heuristic arguments
by Krapivsky \cite{Krapivsky}.

\acknowledgments

The authors are grateful for useful discussions with S.~Cornell,
M.~Droz and M.~Howard.
This work was supported by a grant from the EPSRC, and by NSF grants
DMR 90-07811 and CHE 93-1729.

\appendix

\section*{Calculation of the leading correction term for $\D$}

\begin{figure}
\vskip .2in\centerline{\epsfxsize=3in\epsfbox{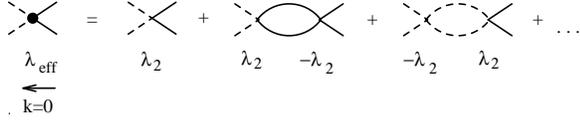}}
\vskip .15in\caption{The expansion for the effective coupling constant.
The wave number integrals are regulated by a cutoff $\L$.}
\label{leff}
\end{figure}

In order to calculate the first order correction term to the expansion
$\D=\D_0+\dots$ we must first comment on the bulk diagrams which generate
$\lf$.  The effective coupling can be calculated as an expansion in the bare
couplings, via the diagrams shown in Fig.~\ref{leff}.  The loop integrals
in this expansion require the cutoff $\L$, and one finds
\begin{equation}\label{newlam}
\lf=\l_2-\l_2^2{4\L^{d-2}\over(8\pi)^{d/2}(d-2)}+O(\l_2^3).
\end{equation}

If the response functions in the loop of Fig.~\ref{surface}(c) were
instead just propagators, then this set of diagrams would be included into
those of Fig.~\ref{surface}(b) when the substitution $\l_2\goto\lf$
is made via (\ref{newlam}). Therefore, the terms which are new and
constitute a correction
to $\D_0$ are those in Fig.~\ref{surface}(c) with the propagator loop
subtracted out.  We define the large $t$ limit of these diagrams to be
$\D_1$, that is
\begin{eqnarray}
\D_1=4\l_2^2\np^2\int_0^\infty &&dt_2\int_0^{t_2}dt_1{d^dk\over(2\pi)^d}
\biggl[{e^{-2k^2t}(1+\np\l_1t_1)^2\over(1+\np\l_1t_2)^4}\nonumber\\
&&-{\L^{d-2}\over(8\pi)^{d/2}(d-2)(1+\np\l_1 t_2)^2}\biggr].
\end{eqnarray}
Performing the wave number integral with the $\L$ cutoff imposed in the same
manner as in (\ref{newlam}) gives
\vbox{
\begin{eqnarray}
&&\D_1={4\l_2^2\np^2\over(8\pi)^{d/2}}\int_0^\infty{dt_2\over
(1+\np\l_1t_2)^4}\nonumber\\\noalign{\medskip}
&&\times\int_0^{t_2}dt_1\biggl[{(1+\np\l_1t_1)^2\over(t_2
-t_1-\L^{-2})^{d/2}} -{\L^{d-2}\over d-2}(1+\np\l_1t_2)^2\biggr].\nonumber\\
\end{eqnarray}}
The $t_1$ integral can be evaluated as a Laplace convolution integral, and
the cutoff dependent terms cancel.  The remaining $t_2$ integral is
\begin{eqnarray}
\D_1={-8\l_2^2\np^2\over(8\pi)^{d/2}(d-2)}&&\int_0^\infty{dt_2\>
t_2^{1-d/2}\over(1+\np\l_1t_2)^4}\nonumber\\
\times&&\biggl[1+{4\np t_2\over 4-d}
+{8\np^2t_2^2\over(4-d)(6-d)}\biggr].
\end{eqnarray}
This integral can be done exactly, giving
\begin{equation}
\D_1={\l_2^2\over\l_1^2}(\np\l_1)^{d/2}{(d+2)(d+4)\over
48(8\pi)^{d/2-1}\sin(\pi d/2)}.
\end{equation}
In terms of the initial density $n_0$ and the effective coupling then
one finds the result (\ref{Delta}) for $\D$.
Evaluating the diagrams such as those in Fig.~\ref{surface}(c), but
containing more loops will then give the higher order terms in this small
$n_0$ expansion of $\D$.

\end{document}